\documentclass[aps,prb,twocolumn,groupedaddress,showpacs,amsmath,amssymb,amsfonts,floatfix]{revtex4}
\usepackage{epsfig}
\usepackage{braket}
\usepackage{psfrag}

\begin{document}

\title{Static versus dynamic fluctuations in the one-dimensional
extended Hubbard model}
\author{H. A. Craig$^{1,2}$, C. N. Varney$^2$, W. E. Pickett$^2$, R. T.
Scalettar$^2$}
\affiliation{$^1$American River College, Sacramento, California 95841,
USA\\
$^2$Physics Department, University of California, Davis,
California 95616, USA}

\begin{abstract}
The extended Hubbard Hamiltonian is a widely accepted model for
uncovering the effects of strong correlations on the phase diagram of
low-dimensional systems, and a variety of theoretical techniques have
been applied to it. In this paper the world-line quantum Monte Carlo
method is used to study spin, charge, and bond order correlations of the
one-dimensional extended Hubbard model in the presence of coupling to
the lattice. A static alternating lattice distortion (the ionic Hubbard
model) leads to enhanced charge density wave correlations at the
expense of antiferromagnetic order.  When the lattice degrees of freedom
are dynamic (the Hubbard-Holstein model), we show that a similar effect
occurs even though the charge asymmetry must arise spontaneously.
Although the evolution of the total energy with lattice coupling is
smooth, the individual components exhibit sharp crossovers at the phase
boundaries.  Finally, we observe a tendency for bond order in the region
between the charge and spin density wave phases.  
\end{abstract}

\maketitle

\section{Introduction}
\label{sec:intro}

The study of strong interaction effects in low-dimensional systems
remains one of the most active fields of research in condensed matter
physics.  The extended Hubbard Hamiltonian (EHH) has been widely
explored as a model of correlation effects in tight-binding systems
and, more specifically, for the competition between different types of
ground state order: charge density wave, antiferromagnetism, and, in
the case of attractive interactions, superconductivity.  In one
dimension, it has also been used to understand the behavior of
materials including conducting polymers\cite{keiss92} and organic
superconductors.\cite{ishiguro90}

The ground state phase diagram of the one-dimensional EHH was first
obtained within a weak coupling renormalization group (RG) calculation.
\cite{emery79,solyom79}  For repulsive on-site interactions $U$ which
are sufficiently large compared to the intersite repulsion $V$,
specifically, for $U > 2 V$, the ground state is a spin density wave
(SDW) phase, with power law decay of spin correlations.  For $2 V> U$,
the ground state has charge density wave (CDW) order.  These charge
correlations exhibit true long range order, that is, they go
asymptotically to a nonzero value at large separations, since the
associated broken symmetry is discrete.  Finally, for attractive
intersite interactions, singlet and triplet superconducting phases
exist at $T=0$, again with power law decays of the associated
correlation functions.

Subsequent to the RG work, the question of the order of the
transitions between these different phases was studied, with a
prediction that for repulsive $U$ and $V$ second-order SDW-CDW
transitions at weak coupling were separated by a tricritical point
from first-order transitions at strong coupling.
\cite{hirsch84,hirsch85,cannon90,cannon91} Up to several years ago,
estimates of the location of the tricritical point varied from
$U_t=1.5t$ to $U_t=5t$ (with $V_t \approx U_t/2$.) More recently, this
picture has been further modified by the suggestion that a narrow
region exhibiting ``bond ordered wave'' (BOW) correlations separates
the SDW and CDW regions at weak coupling. \cite{nakamura99,nakamura00,
sengupta02,tsuchiizu02,sandvik04,tsuchiizu04,tam06}

The competition of CDW and SDW order in the one-dimensional EHH is
further modified if the electrons couple to lattice degrees of
freedom.  In the case where these are static, most investigations have
addressed the case when there is only on-site repulsion $U$, that is,
$V=0$.  In this ``ionic Hubbard model'' the frozen distortions have an
alternating pattern down the chain,\cite{note1} and an additional
issue is the possibility that the band insulator at $U=0$ and half
filling is first driven metallic before becoming a SDW Mott
insulator. \cite{kancharla07,paris07} If the coupling of the electrons
to the lattice is in the form of dynamically varying phonon degrees of
freedom, one has the Hubbard-Holstein or Su-Schrieffer-Heeger
Hamiltonian.

The interplay between band-insulating behavior and electron-electron
interaction effects such as those studied in this paper has recently
been explored in a number of contexts.  Dynamical mean field theory
studies of binary alloy band insulators described by a bimodal
distribution of randomly located one-body potentials have observed
several novel effects, including Mott insulating behavior away from
half filling \cite{byczuk03,byczuk04} and band-insulator to metal
transitions driven by increasing on-site repulsion. \cite{garg06}
Analogous studies of interacting bosons in ``superlattice'' potentials
in which the site energies are modulated have also been used
\cite{jaksch98,buonsante04,buonsante04b,buonsante05,rousseau06} to
describe experiments on ultracold optically trapped (bosonic) atoms.
\cite{thomas02,friebel98,ahmadi05,peil03}

There has been relatively little work, especially using quantum Monte
Carlo (QMC) simulations, which addresses how such lattice coupling
affects the SDW-CDW phase boundary in the EHH in which both $U$ and
$V$ are nonzero.  In this paper, we apply the world-line QMC (WLQMC)
method to the one-dimensional EHH with an additional, static one-body
potential, and with dynamically fluctuating (``Holstein'') phonons.
We quantitatively determine the amount of lattice coupling required to
stabilize a charge ordered phase when the system begins at values of
the electron-electron interactions in the spin density wave regime.
An interesting feature of our results is that the quantum fluctuations
induced by the hopping $t$ have the opposite effect on the strong
coupling ($t=0$) phase boundary in the two cases.  We also present
detailed results for the evolution of the different components of the
energy through the phase transition region.

The remainder of this paper is organized as follows. An explicit
description of our Hamiltonian and a brief review of our numerical
approach are presented in Sec. \ref{sec:model_comp_methods}.  Results
for coupling to static and dynamic lattice deformations are given in
Secs. \ref{sec:results_eihh} and \ref{sec:results_ehhh}, respectively.

\section{Model and Computational Methods}
\label{sec:model_comp_methods}

The extended Hubbard Hamiltonian is
\begin{align}
\widehat H_{\rm el} &= \widehat K + \widehat P,
\nonumber \\
\widehat K &= -t \sum_{i\sigma} (c_{i+1,\sigma}^{\dagger}
c_{i,\sigma} + c_{i,\sigma}^{\dagger} c_{i+1,\sigma} ), \nonumber
\\ 
\widehat P &= U \sum_i n_{i,\uparrow} n_{i,\downarrow}
    +V \sum_i n_{i} n_{i+1}.
\noindent
\end{align}
Here $c_{i,\sigma}^\dagger, c_{i,\sigma}$, and $n_{i,\sigma}$ are the
creation, destruction, and number operators, respectively, for
electrons of spin $\sigma$ at site $i$ of a one-dimensional lattice,
and $n_{i} = n_{i,\uparrow} + n_{i,\downarrow}$.  The hopping $t$
determines the kinetic energy (noninteracting band dispersion
$\epsilon_k = -2 t \, {\rm cos} \, k$), and is set to $t=1$.  $U$ and
$V$, taken to be positive, are the on-site and intersite repulsions.
We will be exclusively interested in the properties of the model at
half filling where the number of fermions $N_f=\sum_i n_i = N$, is
equal to the number of lattice sites.

We will consider additional couplings to an on-site lattice degree of 
freedom,
\begin{align}
\widehat H &= \widehat H_{\rm el} + \widehat H_{\rm lattice},
\nonumber \\
\widehat H_{\rm IHM} &= \Delta \sum_i (-1)^i n_i,
\nonumber \\
\widehat H_{\rm Holstein} &= \lambda \sum_i  x_i n_i
+ \sum_i \left(\frac12 p_i^2 + \frac12 \omega_0^2 x_i^2\right),
\noindent
\end{align}
where $\widehat H_{\rm lattice}$ can take one of two possible forms:
either static (ionic Hubbard model ``IHM'') or dynamic (``Holstein'').
Analytic and numeric studies on such Hamiltonians are quite numerous.
\cite{hubbard81,egami93,ortiz94,resta95,resta99,fabrizio99,wilkens01,
batista04,kancharla07,paris07,zhang02,zhang03,riera00,sil96}

It is useful to review the strong coupling $(t=0)$ phase diagram,
since when the hopping is nonzero the topology of the phase diagram
is rather similar qualitatively and even quantitatively.  In the
absence of an interaction with the lattice, the SDW phase, which
consists of a collection of singly occupied sites, has energy $E_{\rm
SDW}^{t=0} = N V $, while the CDW phase has alternating empty and
doubly occupied sites, and energy $E_{\rm CDW}^{t=0} = NU/2 $.  The
boundary is given by $ V= U/2$.  A static lattice distortion $\Delta$
breaks the twofold symmetry of the CDW state and lowers the energy by
$N \Delta$ on the preferred sublattice.  The resulting boundary is
shifted to $ V=U/2 - \Delta$.

In the case of coupling to a dynamical phonon, we can construct the
$t=0$ phase diagram by completing the square of the electron-phonon
term in the Hamiltonian.  The result is an oscillator with the same
frequency $\omega_0$ and an equilibrium position shifted by $\lambda /
\omega_0^2$.  An attractive on-site interaction $-(\lambda^2 / 2
\omega_0^2) n_{i,\uparrow} n_{i,\downarrow}$ is also generated.  Other
terms can be absorbed into a shifted chemical potential and energy.
As with the static term, the weakening of the on-site $U$ shifts the
strong coupling phase diagram in favor of CDW order.  If $-(\lambda^2
/ 2 \omega_0^2)$ is sufficiently large, pairing correlations can come
to dominate, especially in the doped case.  We will not work in that
parameter regime here.

In order to understand how the quantum fluctuations, which develop as
$t$ increases, modify these simple considerations, we employ the
world-line quantum Monte Carlo (WLQMC) method.  \cite{hirsch82}
Consider first the approach for $\widehat H=\widehat H_{\rm
el}+\widehat H_{\rm IHM}$.  We begin by discretizing the inverse
temperature $\beta$ into intervals $\epsilon = \beta / M$ in the
partition function, and approximating the incremental (imaginary) time
evolution operator by the product of the exponentials of the kinetic
energy and potential energy terms separately.
\begin{align}
Z = {\rm Tr} \, (e^{-\beta \widehat H} ) \approx {\rm Tr}
 \, (e^{-\epsilon \widehat K} e^{-\epsilon (\widehat P + \widehat
 H_{\rm IHM}) } )^M.
\nonumber
\noindent
\end{align}
This Suzuki-Trotter approximation \cite{trotter59,suzuki85}
introduces errors in measurements\cite{fye86,fye87} which are of order
the commutator $[\widehat K,\widehat P]$, that is, $t \, U \,
\epsilon^2$, $t\, \Delta \, \epsilon^2$, and $t \, V \, \epsilon^2$.
Except where otherwise noted, we will choose $\epsilon = 0.25$, which
is sufficiently small that the systematic Trotter errors in the
location of the phase boundary are comparable to those arising from
statistical fluctuations in the Monte Carlo sampling and uncertainties
associated with finite size scaling.

The construction of a path integral for $Z$ is completed by
introducing complete sets of fermion occupation number states $I =
\sum \ket{ n_{i,\sigma} }_\tau \bra{ n_{i,\sigma} }_\tau$ both for the
trace and at all imaginary times, i.e., between each product,
$e^{-\epsilon \widehat K} e^{-\epsilon \left(\widehat P + \widehat
H_{\rm IHM}\right)} $.  The exponentials of the terms in $\widehat P +
\widehat H_{\rm IHM}$ immediately act on the eigenstates, replacing
all operators by numbers.  Thus the weight of a particular occupation
number configuration gets a contribution $W_{\rm P} \, W_{\rm IHM}$,
\begin{align*}
&W_{\rm P}(\{n_{i,\tau,\sigma}\}) = {\rm exp}  \Big( \epsilon
    \sum_{i,\tau} \big[ U n_{i,\tau,\uparrow} n_{i,\tau,\downarrow}
    \\ 
&\hspace{41pt}+\,  V (n_{i,\tau,\uparrow} + n_{i,\tau,\downarrow} )
 (n_{i+1,\tau,\uparrow} + n_{i+1,\tau,\downarrow} \big] \Big),
\\
&W_{\rm IHM}(\{n_{i,\tau,\sigma}\}) = {\rm exp} \Big( \epsilon
  \sum_{i,\tau}  \Delta (-1)^i (n_{i,\tau,\uparrow} +
  n_{i,\tau,\downarrow}) \Big),
\end{align*}
where $\{n_{i,\tau,\sigma}\}$ denotes the space- and
imaginary-time-dependent occupation numbers in the collection of
intermediate states.

\begin{figure}[phtb]
  \includegraphics[width=1.75in,angle=-90]{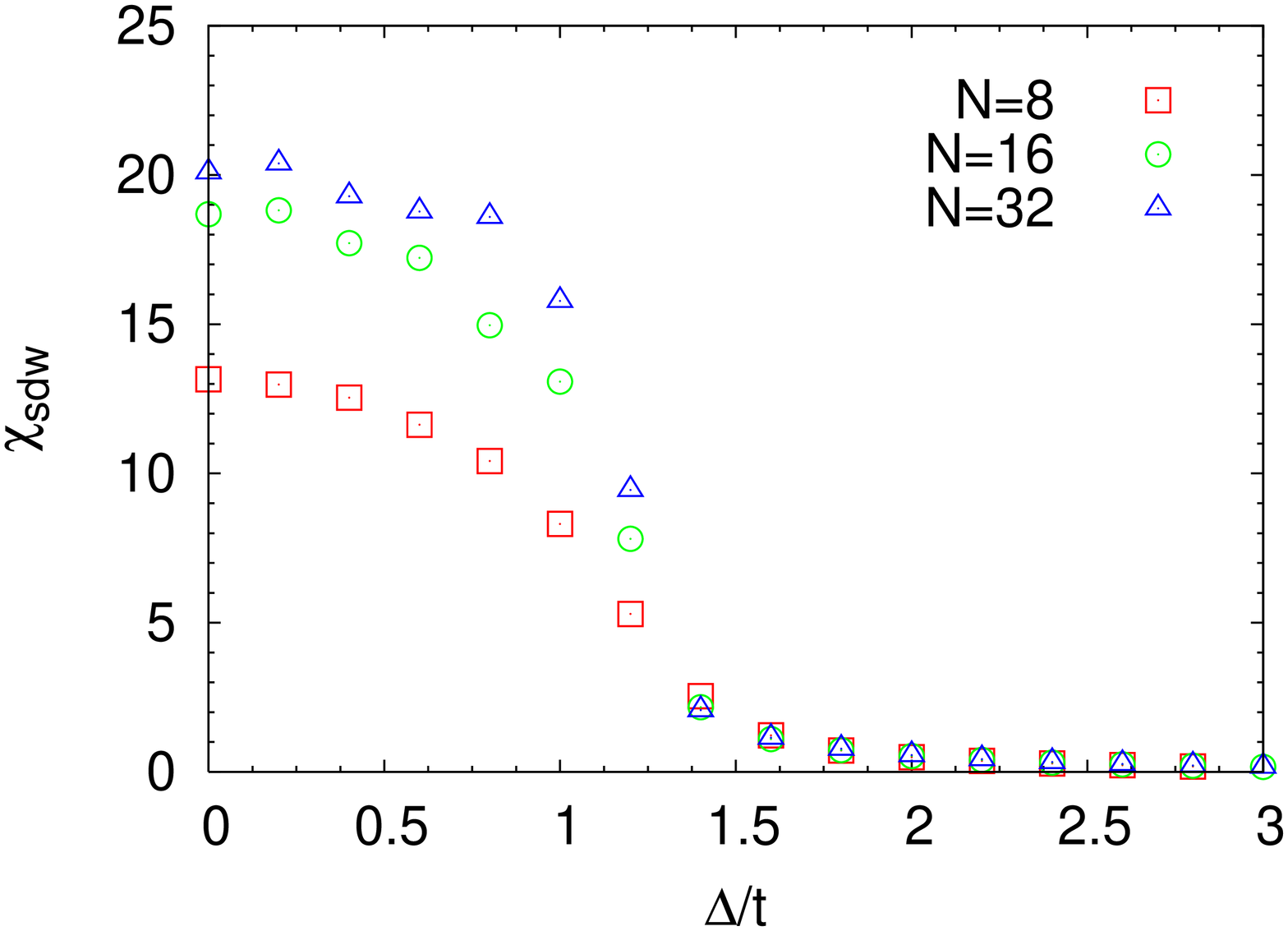}
  \includegraphics[width=1.75in,angle=-90]{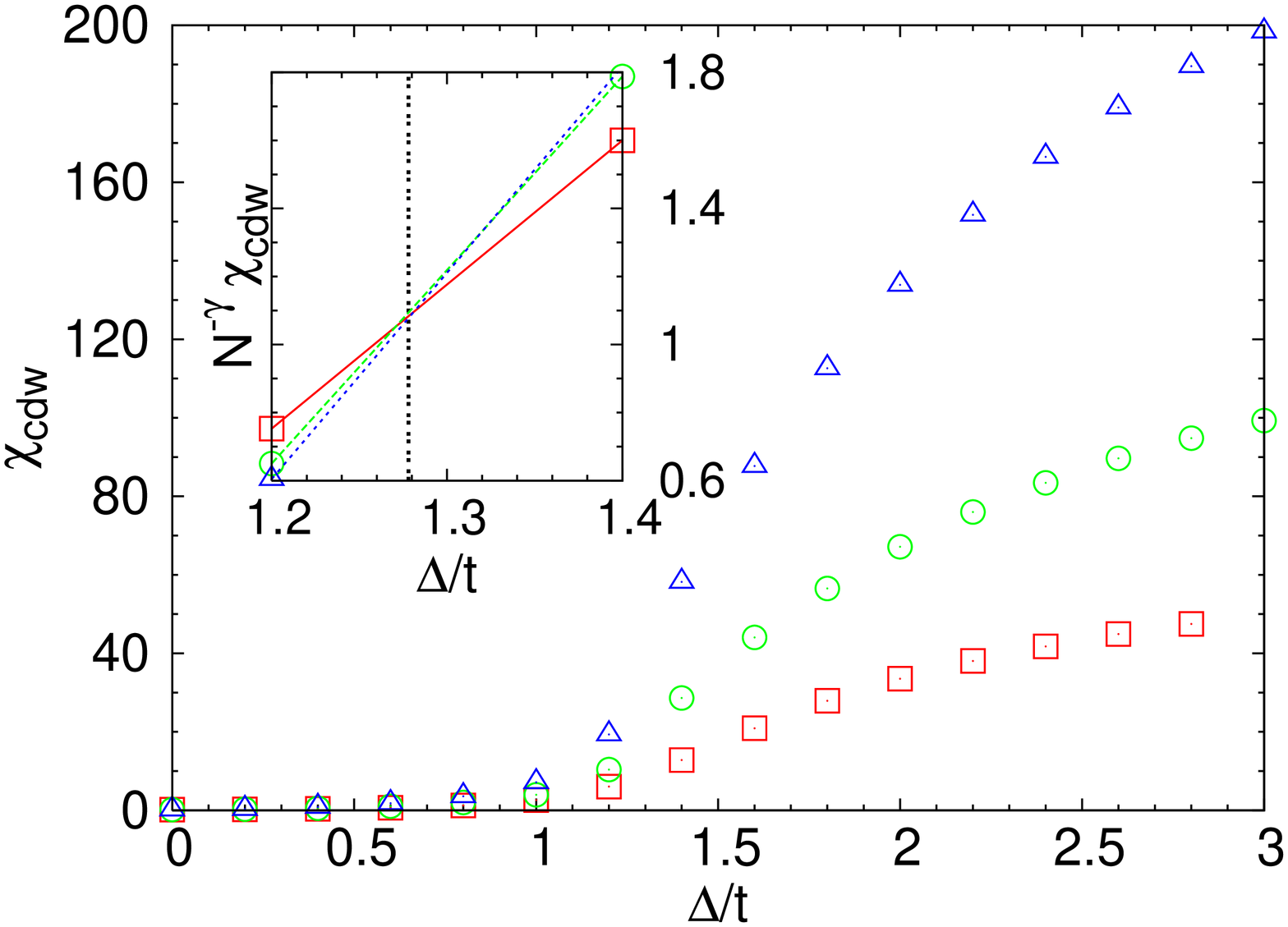}
  \includegraphics[width=1.75in,angle=-90]{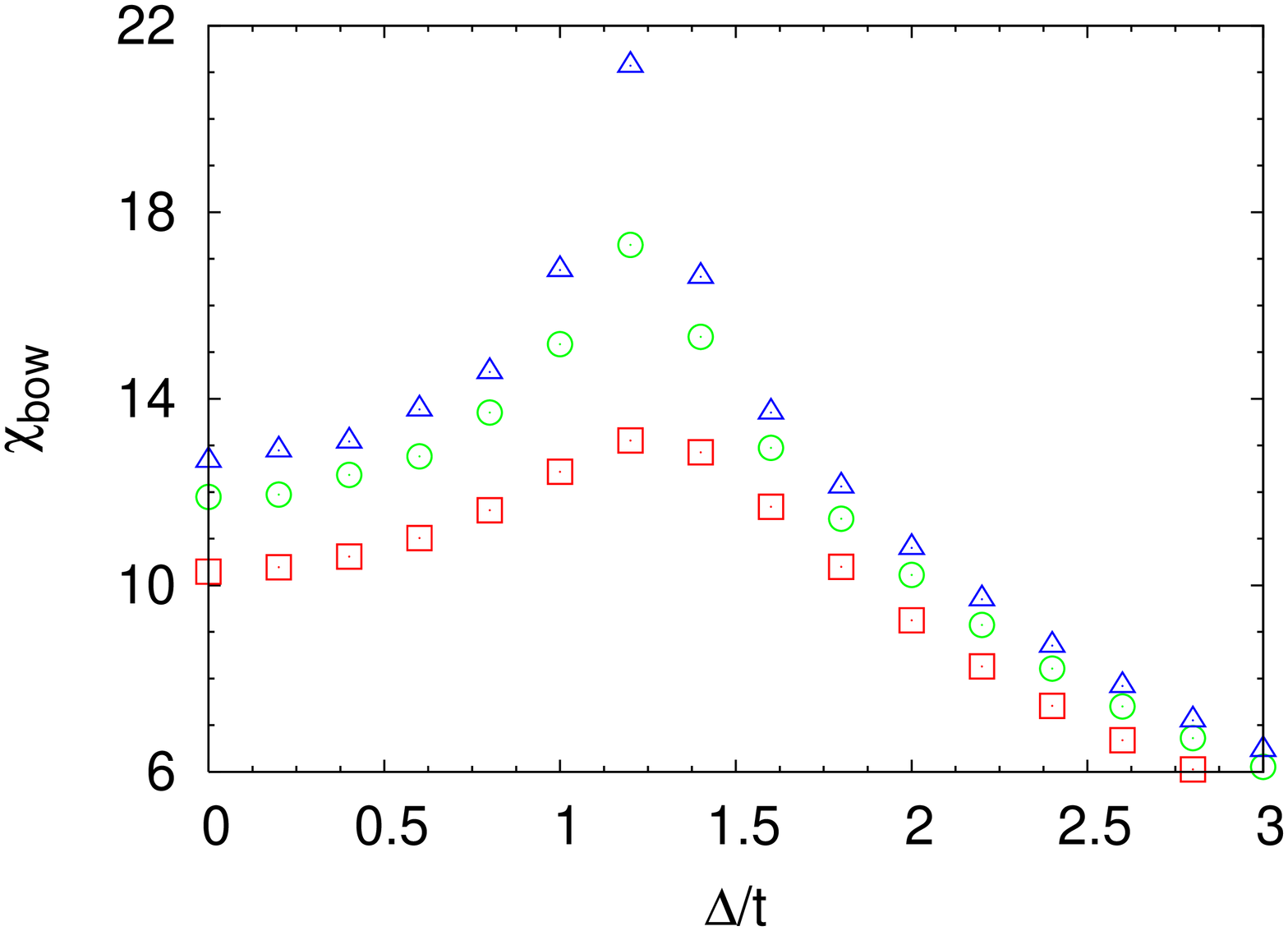}
  \caption{ Spin density wave (top), charge density wave
(middle), and bond ordered wave (bottom) susceptibilities versus
staggered site energy $\Delta$ for $U = 6t$, $V= 1.5t$, $\beta t = 8$,
and $N = 8, 16, 32$.  The SDW-CDW transition occurs at close to the
$t=0$ value, $\Delta=U/2-V$.  BOW correlations are enhanced in the
intermediate region.  In the inset to the central panel, the scaled
$\chi_{\rm CDW}$ is shown for $\gamma = 1$. The scaled
susceptibilities cross at $\Delta_c / t = 1.278$, indicated by the
vertical dotted line.
  \label{fig:stg_unscld_u6v1.5}}
\end{figure}

To accomplish the same replacement of operators by numbers for the
kinetic energy exponentials, $\widehat K$ is further subdivided (the
``checkerboard decomposition'')\cite{hirsch82,barma78} into
\begin{align*}
\widehat K &= \widehat K_{\rm odd}+\widehat K_{\rm even},
\\
\widehat K_{\rm odd} &= -t \sum_{i \,\,{\rm odd},\sigma}
                        (c_{i+1,\sigma}^{\dagger} c_{i,\sigma} +
                        c_{i,\sigma}^{\dagger} c_{i+1,\sigma} ),
\\
\widehat K_{\rm even} &= -t \sum_{i \,\,{\rm even},\sigma}
                        (c_{i+1,\sigma}^{\dagger} c_{i,\sigma} +
                        c_{i,\sigma}^{\dagger} c_{i+1,\sigma} ).
\end{align*}

\begin{figure}[phtb]
  \psfrag{i-2}{\quad\scriptsize{$i-2$}}
  \psfrag{i-1}{\quad\scriptsize{$i-1$}}
  \psfrag{i}{\scriptsize{$i$}}
  \psfrag{i+1}{\quad\scriptsize{$i+1$}}
  \psfrag{i+2}{\quad\scriptsize{$i+2$}}
  \psfrag{i+3}{\quad\scriptsize{$i+3$}}
  \includegraphics[width=\columnwidth]{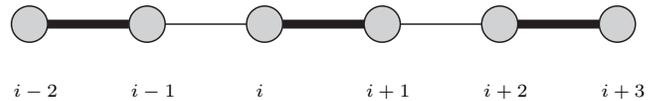}
  \caption{Illustration of a BOW phase. The thick lines indicate a
    high kinetic energy while the thin lines indicate a small kinetic
    energy.
    \label{fig:bow}}
\end{figure}

The expectation value of $\widehat K_{\rm odd}$ and $\widehat K_{\rm
even}$ between the occupation number states $\ket{ n_{i,\sigma}
}_\tau$ and $\bra{ n_{i,\sigma} }_{\tau+1}$ then reduces to a product
of independent two-site problems which can be solved analytically.
Since particle number is conserved in each hopping process, the number
of electrons on each pair of sites in the two states to the left and
to the right of the exponential is identical.  Thus the world lines
generated by connecting all occupied sites ($n_{i,\tau,\sigma}=1$) are
continuous.  The four nonzero matrix elements are
\begin{align*}
\braket{ \, 00 \, | \, e^{\epsilon t (c_1^\dagger c_2 +
c_2^\dagger c_1) } \, | \, 0 0 \, } &= 1,\\
\braket{ \, 11 \, | e^{\epsilon t (c_1^\dagger c_2 +
c_2^\dagger c_1) } \, | \, 1 1 \, } &= 1,\\
\braket{ \, 1 0 \, | \, e^{\epsilon t (c_1^\dagger c_2 +
c_2^\dagger c_1) } \, | \, 1 0 \, } &= \cosh(t \epsilon),\\
\braket{ \, 1 0 \, | \, e^{\epsilon t (c_1^\dagger c_2 +
c_2^\dagger c_1) } \, | \, 0 1 \, } &= \sinh(t \epsilon).
\end{align*}
The product of all these factors over the space-time lattice
constitutes a second contribution $W_{\rm K}$ to the weight associated
with the configuration. Thus, the total weight is $W_{\rm tot} =
W_{\rm P} W_{\rm IHM} W_{\rm K}$. Because all of the matrix elements
are positive in one dimension, the WLQMC algorithm does not exhibit a
sign problem.

In the case $\widehat H = \widehat H_{\rm el} + \widehat H_{\rm
Holstein}$, the trace and intermediate states include not only fermion
occupation labels, but also a complete set of phonon position
eigenstates.  As with $\widehat H_{\rm el}$, the exponential of the
phonon kinetic and potential energies is discretized and split apart.
The result is that in addition to the electronic contributions $W_P \,
W_K$ there is a final phonon piece,
\begin{align*}
W_{\rm ph}(\{x_{i,\tau}\}) =
{\rm exp}  \left[ \frac{1}{2} \epsilon \sum_{i,\tau} \omega_0^2 
x_{i,\tau}^2 +
\left(
\frac{x_{i,\tau+1}-x_{i,\tau}}{\epsilon}
\right)^2
\right].
\end{align*}

Let us then summarize the basic features of the simulation.  The
degrees of freedom being summed over are two space-time arrays of
occupation numbers $n_{i,\tau,\uparrow}$ and $n_{i,\tau,\downarrow}$,
and, in the Holstein case, a space-time array of phonon coordinates
$x_{i,\tau}$, with $i=1,2,\ldots,N$ and $\tau=1,2,\ldots, 2M$.  (The
factor of 2 comes from the checkerboard decomposition.) The total
weight of the configuration is $W_{\rm tot} = W_{\rm P} W_{\rm K}
W_{\rm Ph}$.  The elemental Monte Carlo moves consist of local
distortions of the continuous world lines, together with updates of
the phonon degrees of freedom.  Moves are accepted or rejected
according to the Metropolis algorithm: a random number $0< r < 1 $ is
generated and the move is accepted if $r < W_{\rm tot}^\prime/W_{\rm
tot}$.

The WLQMC algorithm can suffer from long autocorrelation times.  Other
approaches such as the stochastic series expansion method
\cite{sandvik91,sandvik92, sandvik99} and loop algorithms
\cite{prokofev98} can be used to speed up the evolution in phase
space.  Here we confine ourselves only to introducing global
moves\cite{global} in the phonon degrees of freedom to address even
more serious large autocorrelation times there.

\begin{figure}[tb]
  \includegraphics[width=1.75in,angle=-90]{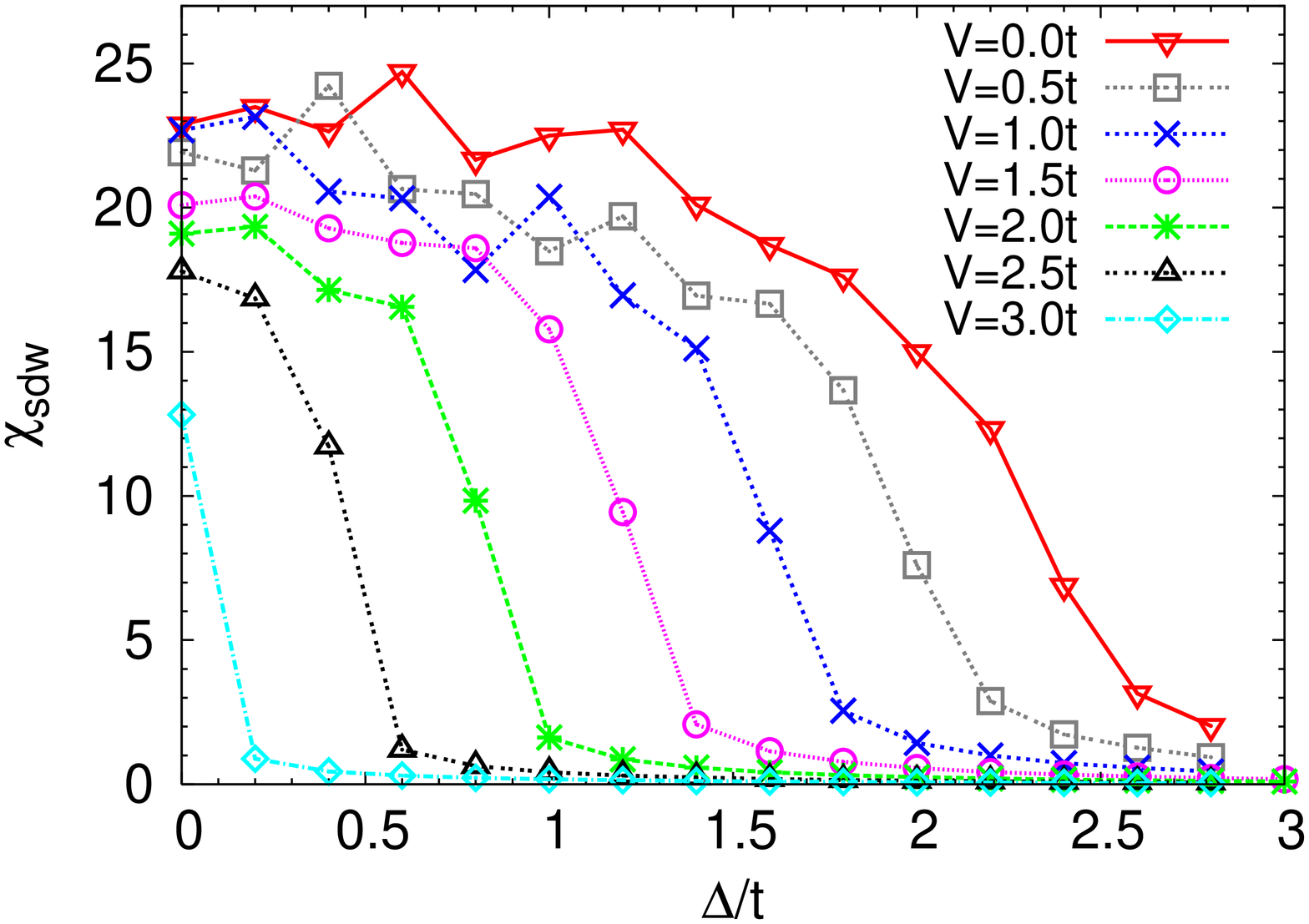}
  \includegraphics[width=1.75in,angle=-90]{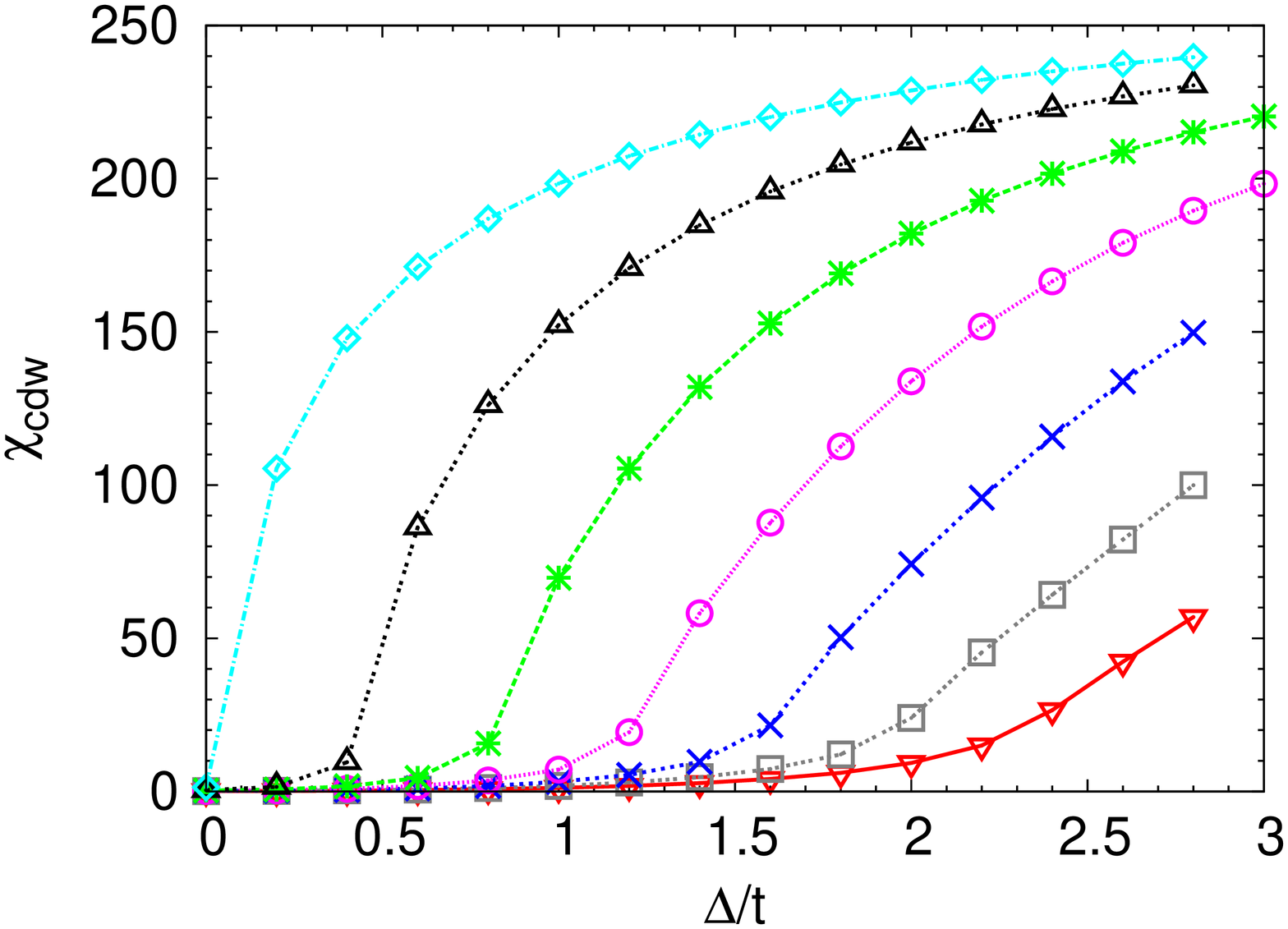}
  \includegraphics[width=1.75in,angle=-90]{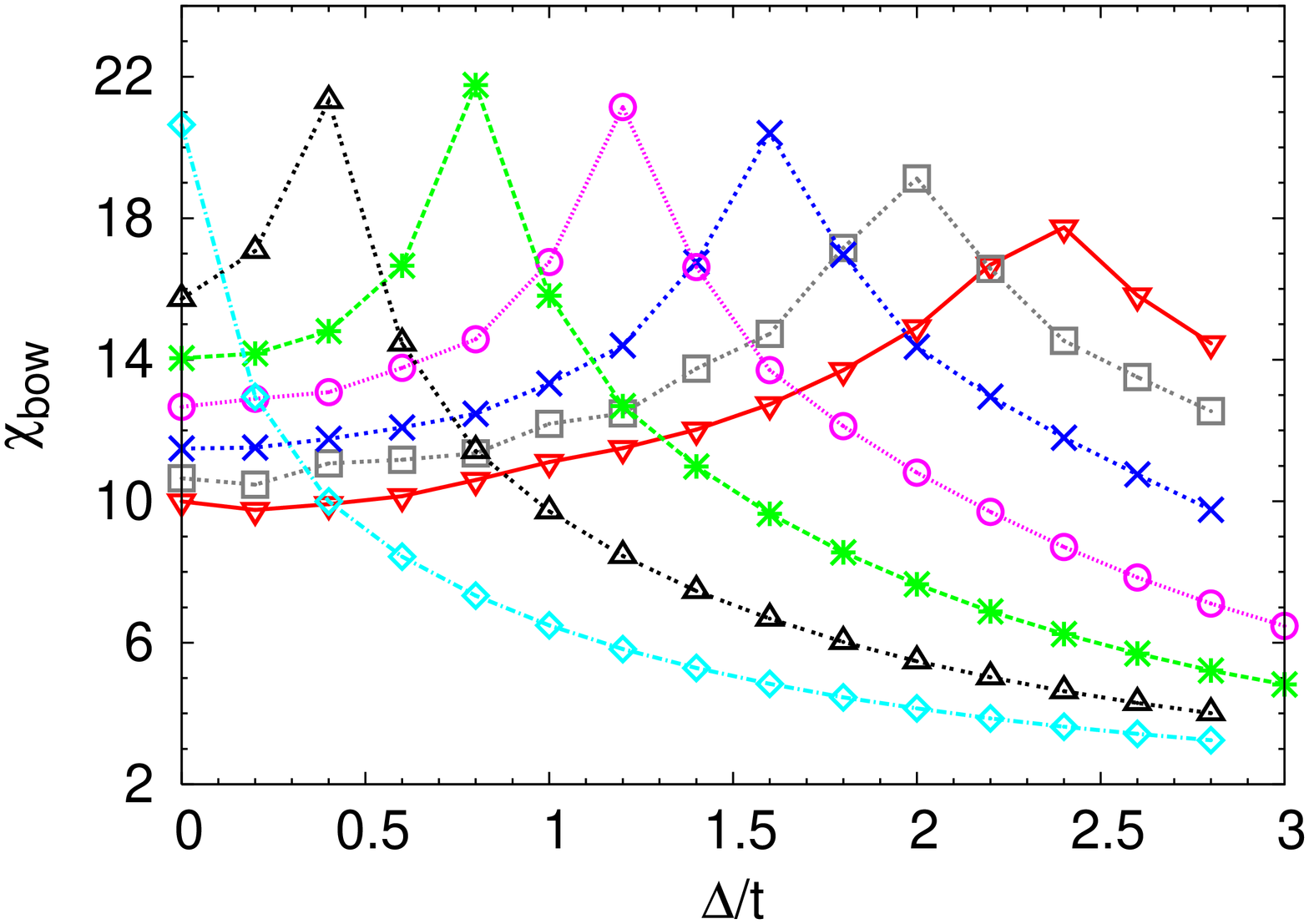}
  \caption{ Spin density wave (top), charge density wave (middle), and
bond ordered wave (bottom) susceptibilities versus staggered site
energy $\Delta$ for $U = 6t$, $V=0.0t, (0.5t), 3.0t$, and $N = 32$.
  \label{fig:stg_susc}}
\end{figure}

\begin{figure}[phtb]
  \includegraphics[width=1.96in,angle=-90]{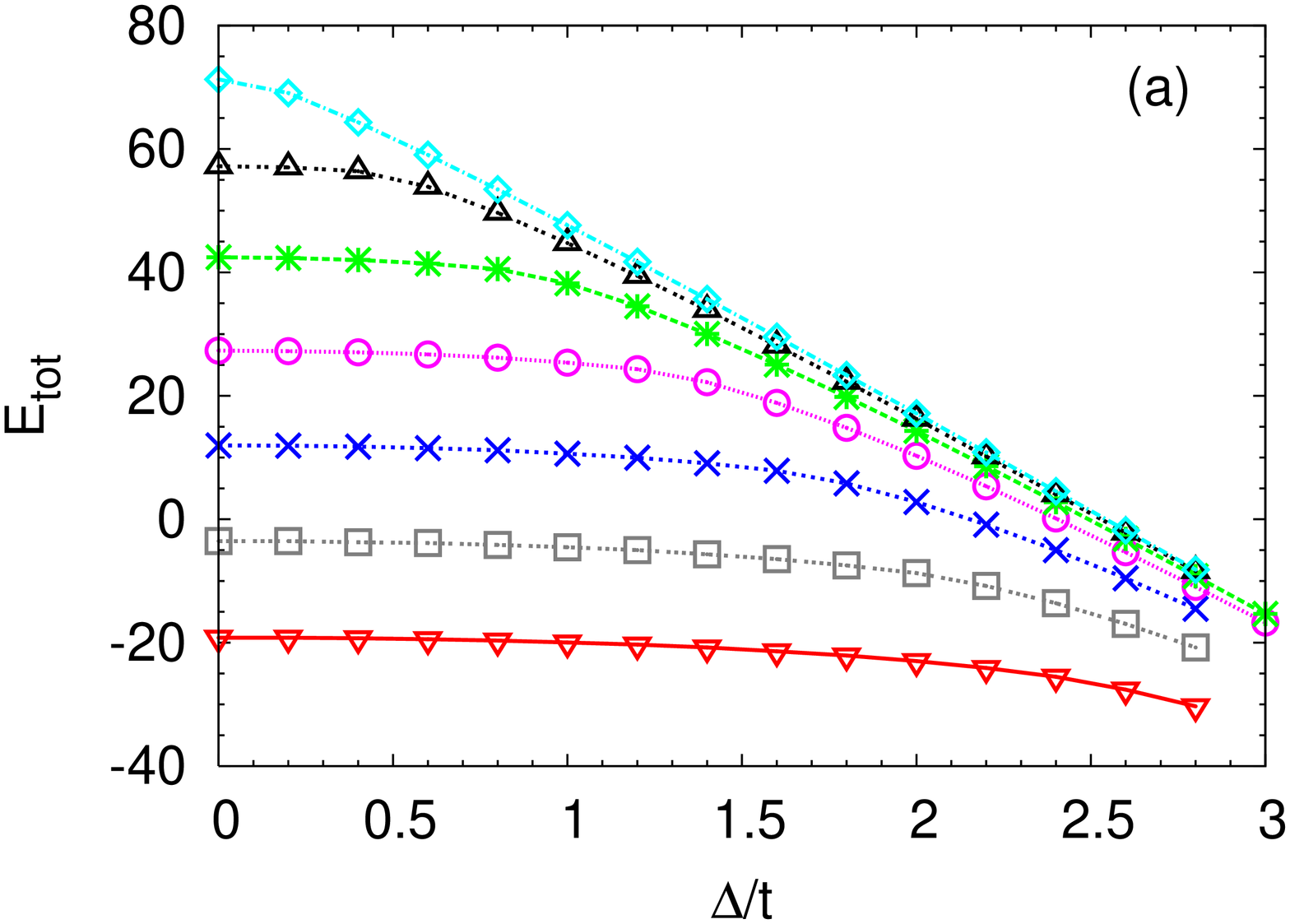}
  \includegraphics[width=1.96in,angle=-90]{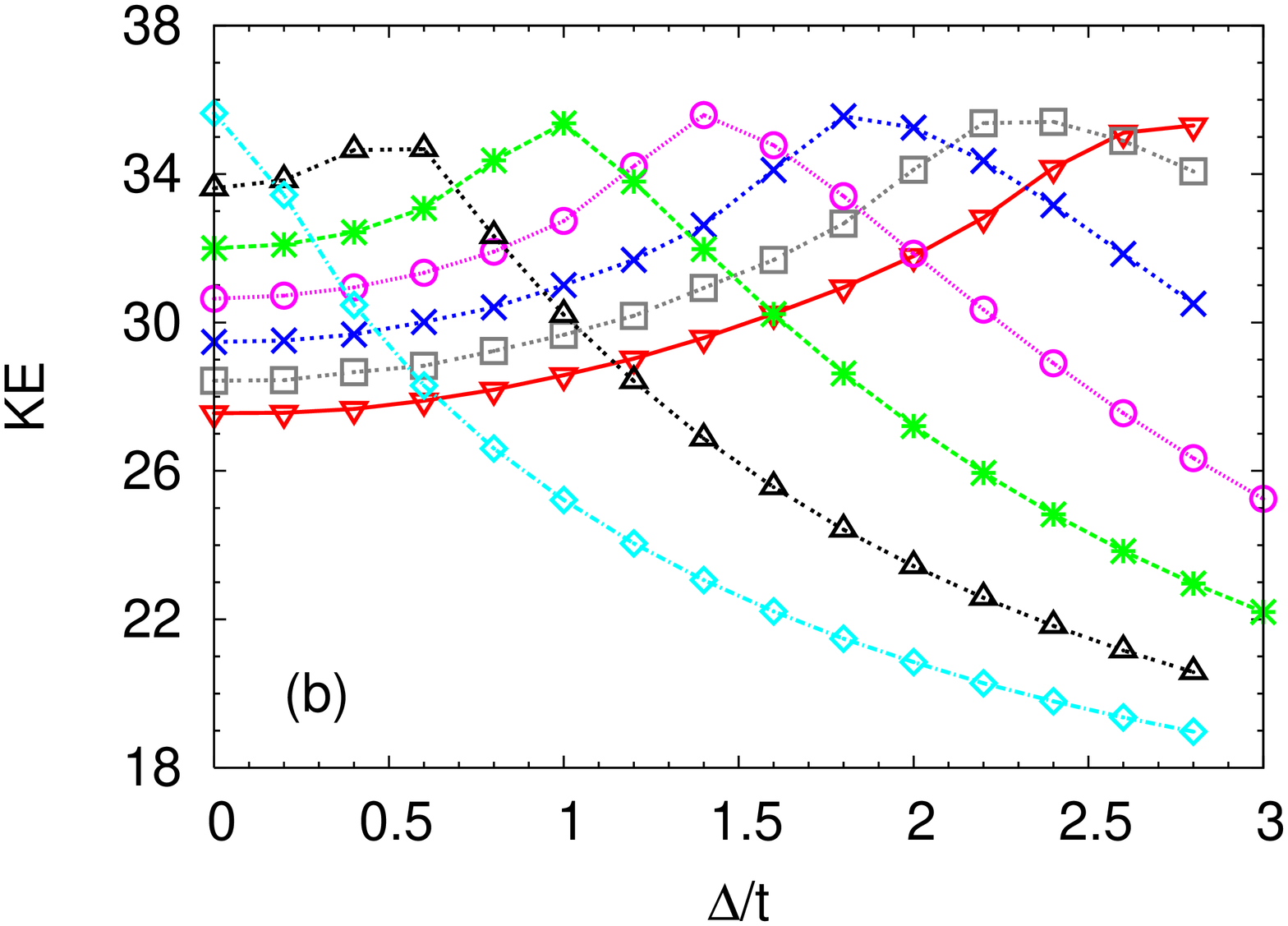}
  \includegraphics[width=1.96in,angle=-90]{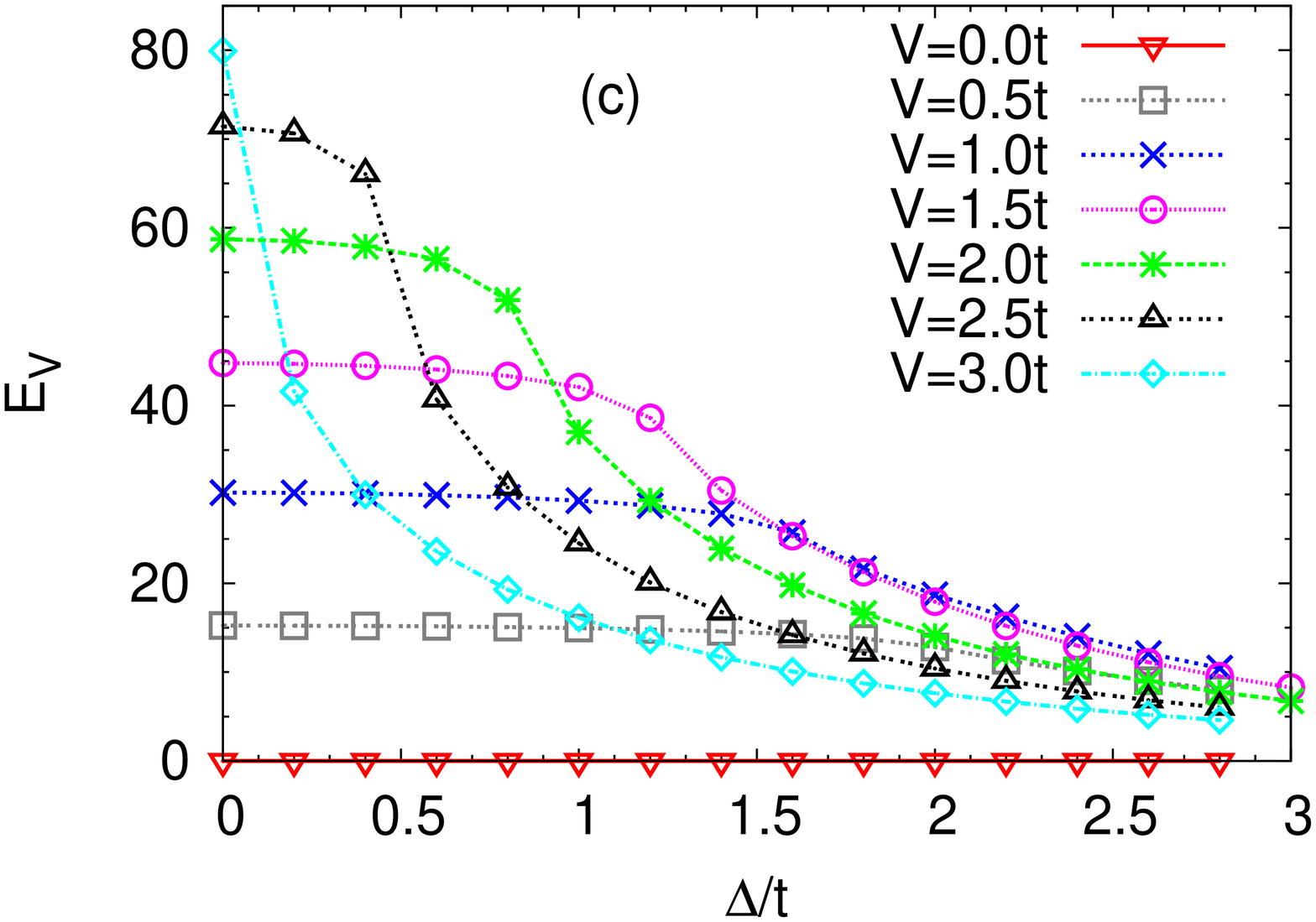}
  \includegraphics[width=1.96in,angle=-90]{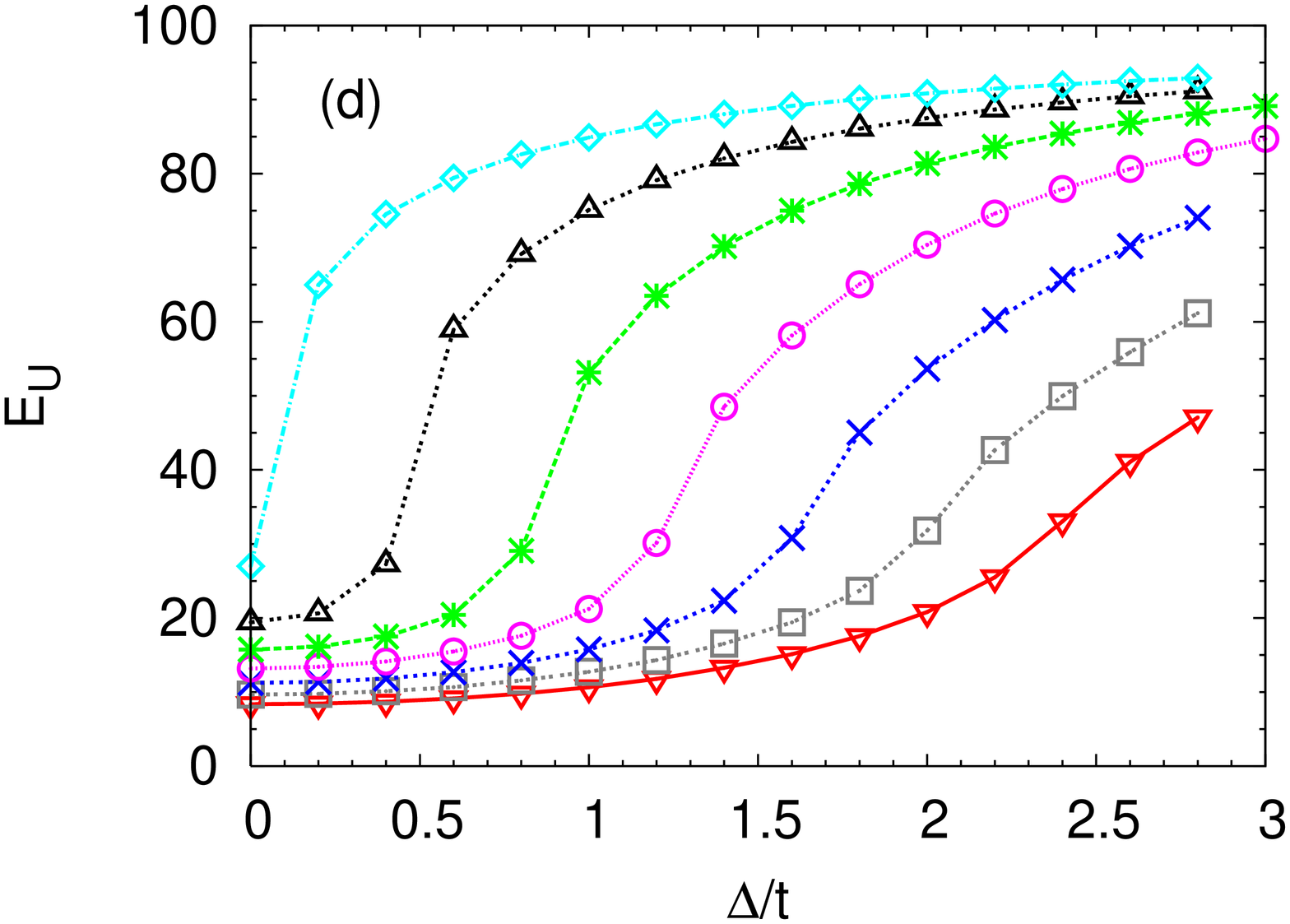}
  \caption{ Total (a), kinetic (b), intersite Coulomb (c), and on-site
Coulomb (d) energies at fixed $U = 6t$ and different $V$. The inverse
temperature is fixed at $\beta t = 8$ and the lattice size at $N =
32$. The kinetic energy is largest in the region where there is a
balance between the CDW and SDW insulating tendencies, in good
correspondence with the behavior of $\chi_{\rm BOW}$.  The intersite
interaction energy falls abruptly on entry to the CDW state, and the
on-site energies rise steeply as the pairs form.
  \label{fig:stg_en}}
\end{figure}

We conclude with a discussion of the observables we will measure.  The
various components of the energy exhibit sharp features as the phase
boundaries are crossed.  Real space spin, charge (relative to the
mean), and bond operators are defined by
\begin{align*}
m(l,\tau) &=
    n_{l,\tau,\uparrow} - n_{l,\tau,\downarrow},
\\
n(l,\tau) &= 
 n_{l,\tau,\uparrow} + n_{l,\tau,\downarrow}-1,
\\
k(l,\tau) &= 
   \sum_{\sigma} 
 (c_{l+1,\sigma}^\dagger(\tau) c_{l,\sigma}(\tau)
 + c_{l,\sigma}^\dagger(\tau) c_{l+1,\sigma}(\tau)).
\end{align*}

The associated correlation functions are
\begin{align*}
c_{\rm spin}(l,\tau) &= \braket{ m(l,\tau) m(0,0) }, \\
c_{\rm charge}(l,\tau) &= \braket{ n(l,\tau) n(0,0) }, \\
c_{\rm bond}(l,\tau) &= \braket{ k(l,\tau) k(0,0) },
\end{align*}
where $(0,0)$ is some reference site in our system.  The local moment
is defined as $\braket{ m_z^2 } = c_{\rm spin}(0,0)$.

\begin{figure}[tb]
  \includegraphics[width=1.85in,angle=-90]{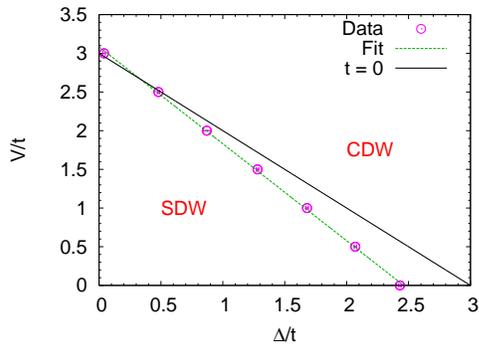}
  \caption{Phase diagram in the intersite $V$ and staggered site
  energy $\Delta$ plane, with $U = 6t$ and $\beta t = 8$.  Line with
  symbols is the result of the WLQMC simulations in this paper.  We
  also show the exact result (line without symbols) for $t = 0$.  As
  expected, the strong coupling limit works well at large $V$, but
  there are significant deviations as $V$ becomes smaller.
  \label{fig:stg_pd}}
\end{figure}

We will also look at the Fourier transforms of these quantities.  The
equal time spin structure factor is
\begin{align*}
S_{\rm spin}(q)
&= \frac{1}{N} \sum_l e^{iql} c_{\rm spin}(l,0),
\end{align*}
with analogous definitions for $S_{\rm charge}$ and $S_{\rm bond}$.
The corresponding zero-frequency susceptibility is
\begin{align*}
\chi_{\rm spin}(q)
&= \frac{1}{N} \sum_{\tau} \sum_l e^{iql} c_{\rm spin}(l,\tau),
\end{align*}
again with analogous definitions for $\chi_{\rm charge}$ and
$\chi_{\rm bond}$.

At half filling the largest responses in the Hubbard model are at
wave vector $q =\pi$.  In a disordered phase, $c(l,0)$ decays
exponentially to zero with the site separation $l$, and the structure
factor is independent of lattice size $N$.  If true long range order
develops, then the structure factor grows linearly with lattice size,
with the factor $e^{i \pi l}$ providing the necessary phases so that
the oscillating $c(l)$ add constructively.  The susceptibility
similarly examines the asymptotics in imaginary time, diverging with
$\beta$ when $c(l,\tau)$ remains nonzero for large $\tau$.

\section{Results: Extended Ionic Hubbard Hamiltonian}
\label{sec:results_eihh}

In the extended Hubbard Hamiltonian with $U=6t$ and $V=1.5t$, we are
well within the SDW phase since $U > 2V$.  In
Fig.~\ref{fig:stg_unscld_u6v1.5} we see that as $\Delta$ is increased,
the SDW susceptibility decreases and the CDW susceptibility grows.
Indeed, $\chi_{\rm CDW}$ rises dramatically in the vicinity of
$\Delta=U/2-V$, as suggested by the strong coupling analysis.  The
transition becomes increasingly sharp as the lattice size is
increased.  Because the CDW correlations break a discrete symmetry,
true long range order is possible at $T=0$.  With our normalization
conventions we expect the CDW structure factor and susceptibility to
grow linearly with lattice size after the onset of long range order.
This is borne out in the central panel of
Fig.~\ref{fig:stg_unscld_u6v1.5}.  The inset in this panel shows a
scaled version of the raw data for $\chi_{\rm CDW}$.  A crossing of
the curves for different lattice sizes $N$ allows us to determine the
location of the critical point.

\begin{figure}[phtb]
  \includegraphics[width=1.85in,angle=-90]{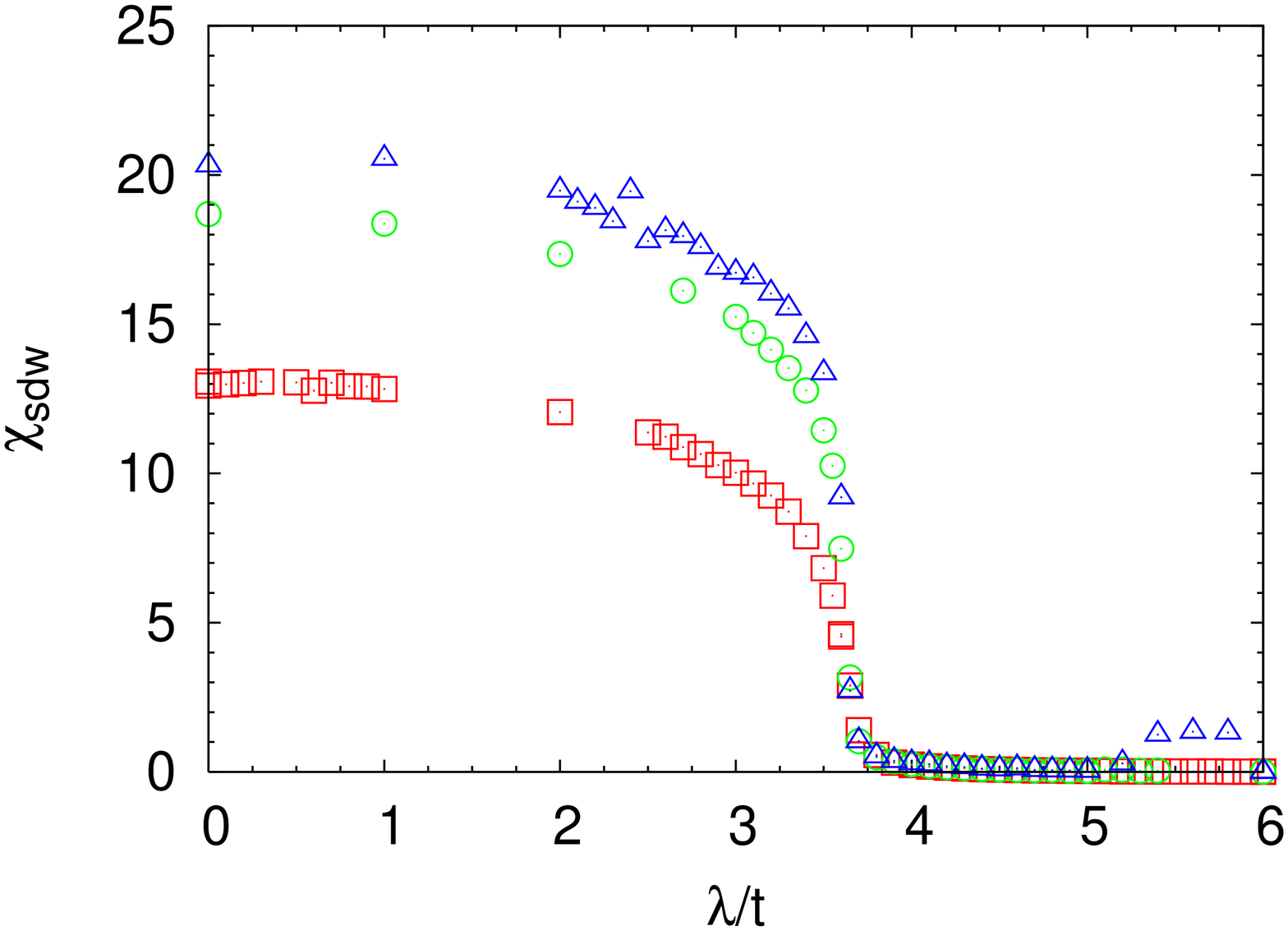}
  \includegraphics[width=1.85in,angle=-90]{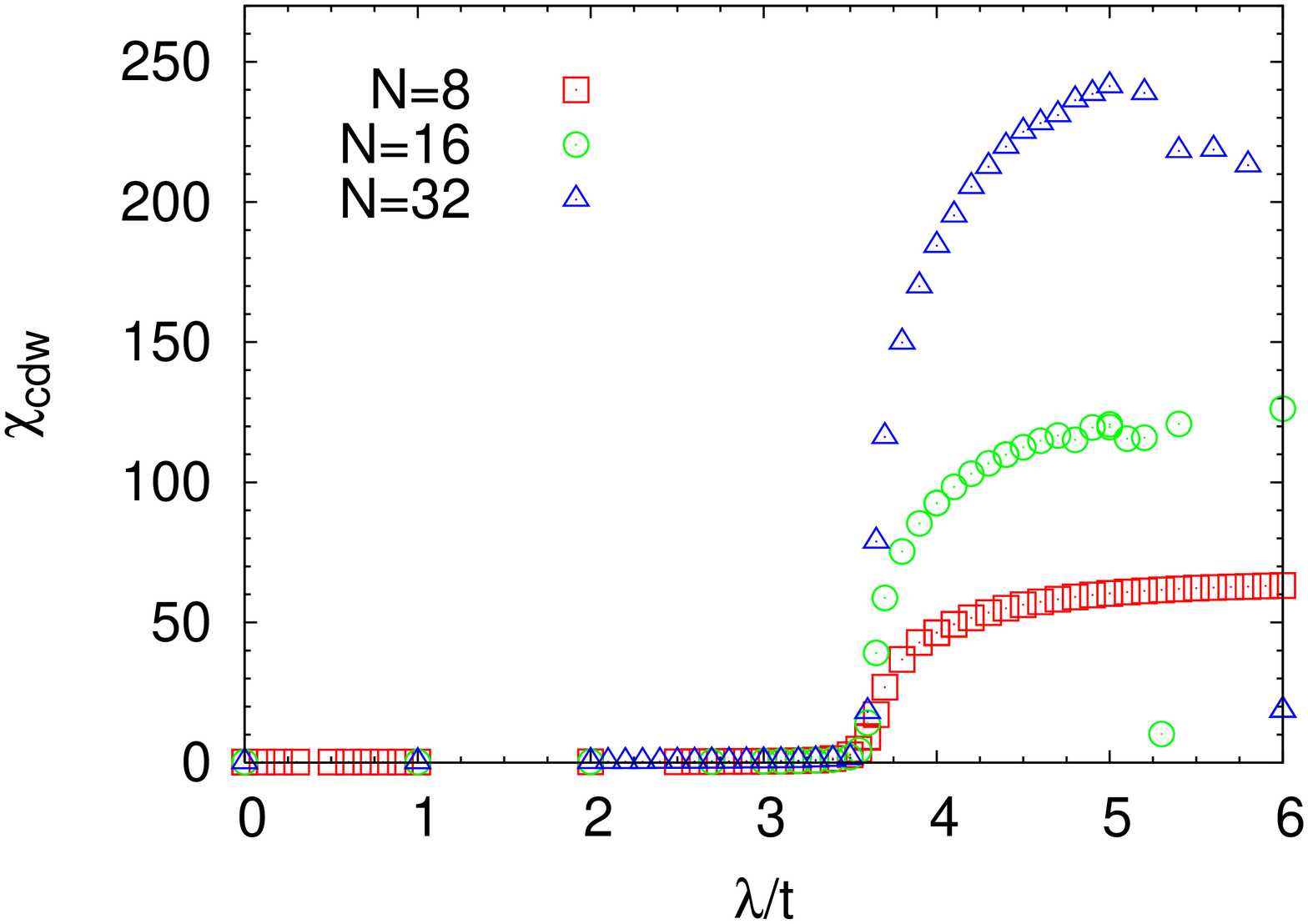}
  \includegraphics[width=1.85in,angle=-90]{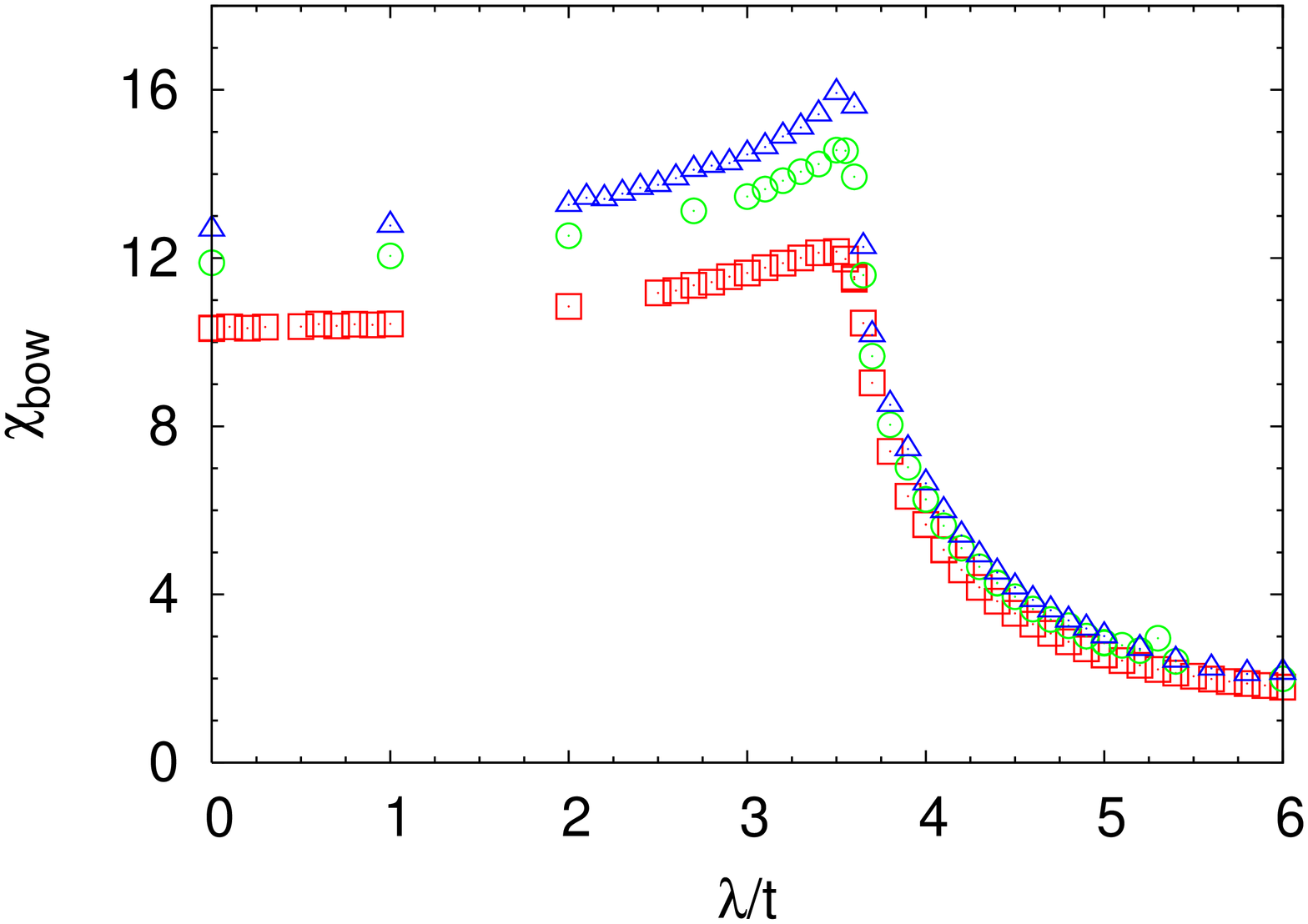}
  \caption{Spin density wave (top), charge density wave (middle), and
  bond ordered wave (bottom) susceptibilities versus electron-phonon
  coupling $\lambda$ for $U = 6t$, $V= 1.5t$, $\omega_0 = 2t$, $\beta
  t = 8$, and $N = 8, 16, 32$.
  \label{fig:phn_unscld_u6v1.5}}
\end{figure}

The SDW correlations that are dominant at small $\Delta$ break a
continuous symmetry, and hence in one dimension decay with a power law
at $T=0$, that is, $c_{\rm spin}(l,0) \propto 1/l$.  This behavior
accounts for the relatively less rapid growth of the SDW
susceptibility with lattice size.

It is important to make another distinction between the CDW and SDW
phases, the phases that arise as broken symmetries from the
interaction terms $V$ and $U$, and the staggered density which is
caused by the one-body term $\Delta$.  This staggered potential
$\Delta$ breaks translational invariance so that there is a small
degree of CDW order even in the SDW phase.  By contrast, in a
competition solely between $U$ and $V$ at $\Delta=0$, no CDW order
would exist in the SDW phase.

The bottom panel of Fig.~\ref{fig:stg_unscld_u6v1.5} shows the BOW
correlations.  In a BOW phase the kinetic energy on the links
oscillates between two values as one traverses the chain (see
Fig.~\ref{fig:bow}).  SDW correlations are immediately plausible after
observing that $U$ leads to singly occupied sites (moment formation)
and that neighboring spins that are antiparallel have a second-order
lowering of their energy ($\Delta E^{(2)} \propto -t^2/U$) relative to
neighboring spins that are parallel.  Analogous reasoning applies to
CDW correlations.  A picture of the less familiar BOW order is the
following: consider a CDW pattern of doubly occupied and empty sites.
A fermion hopping from doubly occupied site $i$ onto neighboring empty
site $i+1$ will prevent, through the interaction $U$, the hopping of a
second electron from doubly occupied site $i+2$ onto $i+1$.  Instead,
an electron on site $i+2$ would prefer to hop to $i+3$.  Thus the
bonds $(i,i+1)$ and $(i+2,i+3)$ have high kinetic energy, while the
intermediate bond $(i+1,i+2)$ has low kinetic energy.  This way of
understanding the origin of BO invokes both CDW and SDW correlations,
making it plausible that the BOW might form on the boundary between
the two.

Since the BOW phase also breaks a discrete translational symmetry, the
associated ground state order should be long ranged.  As mentioned in
the Introduction, in the extended Hubbard model ($\Delta=0$) the
original picture of the phase diagram was one with only SDW and CDW
regions, with a weak coupling second-order transition changing at a
tricritical point to a strong coupling first-order transition.
\cite{hirsch84,hirsch85,cannon90,cannon91} Recent QMC simulations with
the stochastic series expansion (SSE) have suggested instead that, at
weak coupling, as $V$ is increased at fixed $U$ there are two separate
transitions: a SDW-BOW transition of the Kosterlitz-Thouless type,
followed by a second-order BOW-CDW transition.  These transitions
merge at a multicritical point into a single, direct, first order
SDW-CDW transition line at strong coupling. \cite{sengupta02}  The
multicritical point was found to be at $(U_m,V_m)=(4.7\pm 0.1,2.51
\pm 0.04)$.

Aspects of this conclusion had been challenged by density matrix
renormalization group calculations \cite{jeckelmann02,sandvik03,
jeckelmann03}.  In particular, the suggestion is that the BOW phase
exists only precisely on the SDW-CDW transition line, as opposed to
being present in an extended region.  Moreover, rather than starting
at $U=V=0$ and reaching out to the multi-critical point, the BOW line
was concluded to begin at finite, nonzero coupling and also extend
somewhat beyond the numerical value for the multi-critical point
obtained using the SSE.  Further SSE calculations \cite{sandvik04} and
functional RG treatments \cite{tam06} appear to confirm earlier SSE
work.

\begin{figure}[t]
  \includegraphics[width=1.75in,angle=-90]{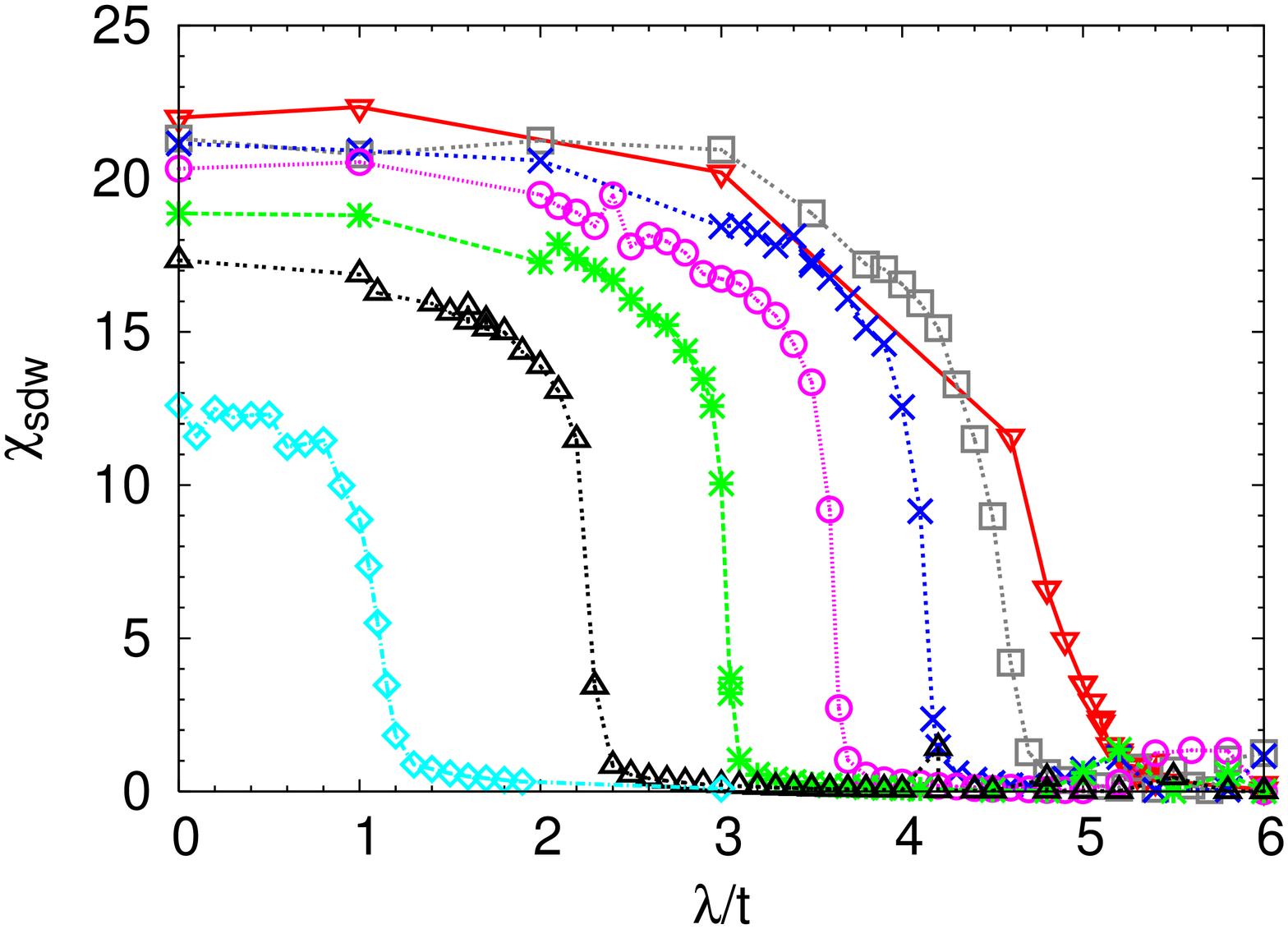}
  \includegraphics[width=1.75in,angle=-90]{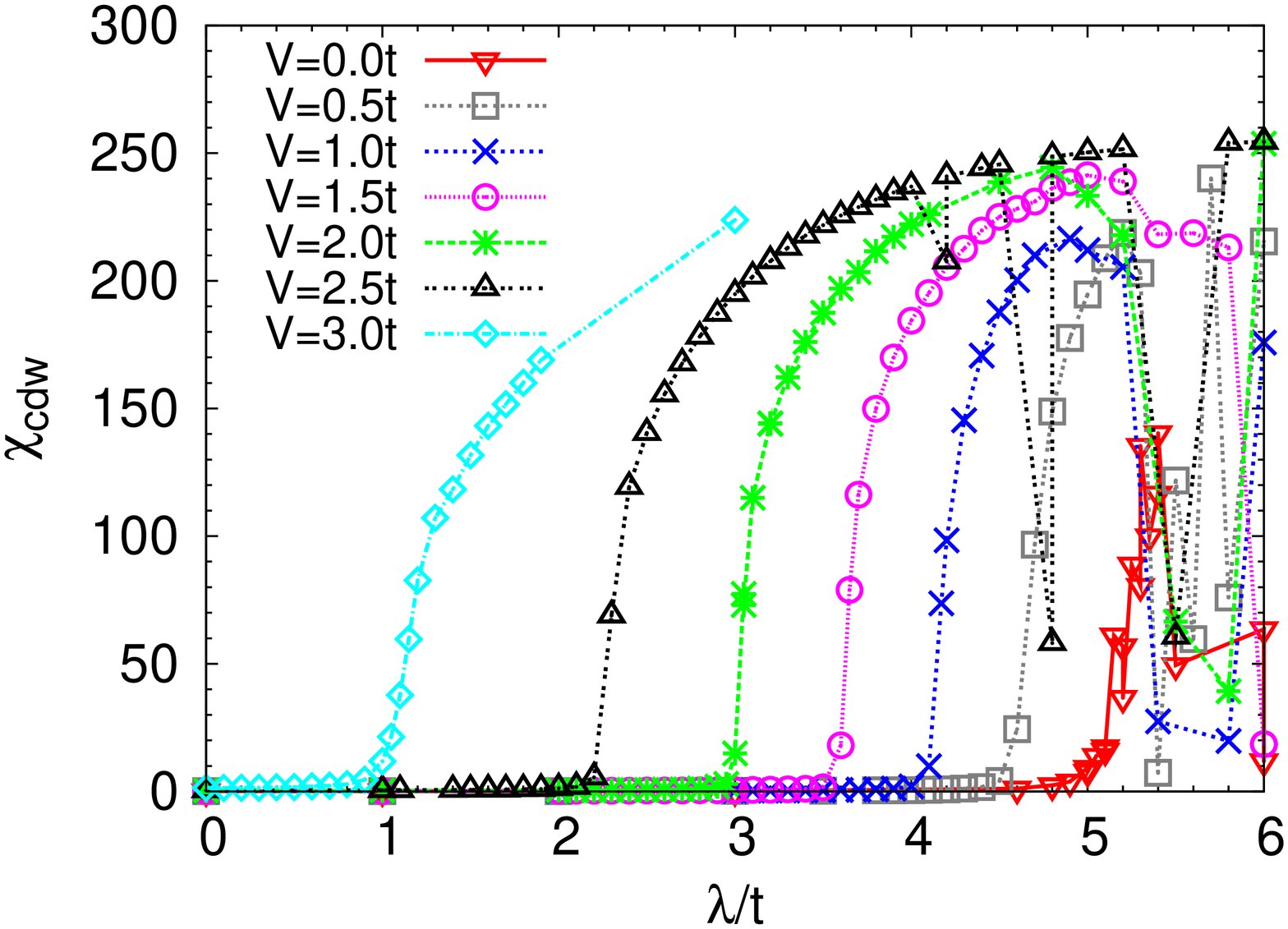}
  \includegraphics[width=1.75in,angle=-90]{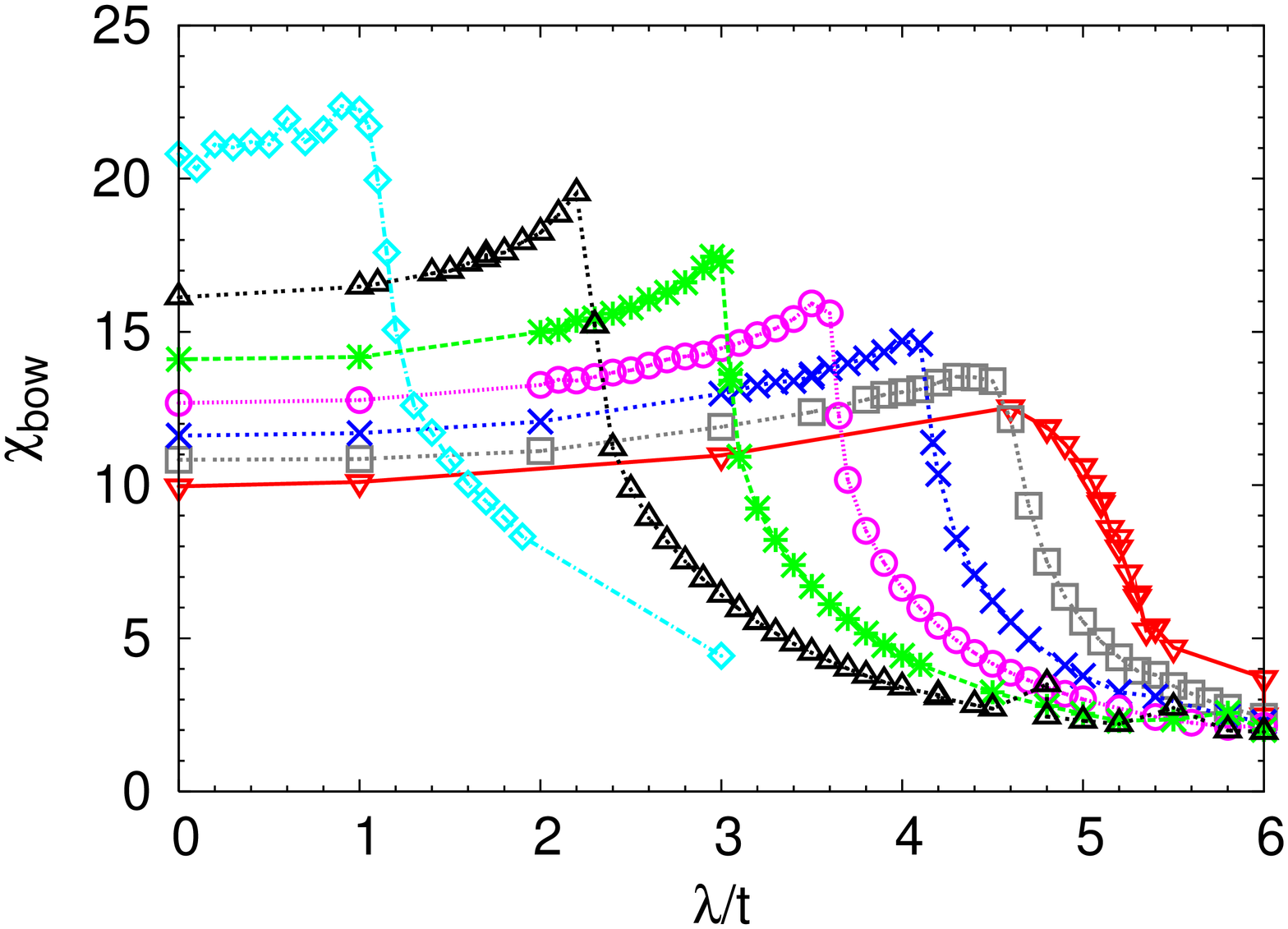}
  \caption{Spin density wave (top), charge density wave (middle), and
  bond ordered wave (bottom) susceptibilities versus electron-phonon
  coupling $\lambda$ for $U = 6t$, $V=0.0t, (0.5t), 3.0t$, $\omega_0 =
  2t$, and $N = 32$.
  \label{fig:phn_susc}}
\end{figure}

We do not propose here to add to this discussion, since our main focus
is on the shift in the SDW-CDW phase boundary.  Indeed, the value of
$U$ in Fig.~\ref{fig:stg_unscld_u6v1.5} is large enough that we would
likely be above the BOW region of the phase diagram.  Nevertheless,
the bottom panel of Fig.~\ref{fig:stg_unscld_u6v1.5} does indicate a
pronounced maximum in $\chi_{\rm BOW}$ near the SDW-CDW transition,
hinting that such order may be present at weaker coupling.  If long
range BO were to exist, we would expect to see $\chi_{\rm BOW}$ grow
linearly with $N$, as does $\chi_{\rm CDW}$.  This is clearly not the
case for the parameters and lattice sizes of
Fig.~\ref{fig:stg_unscld_u6v1.5}.

\begin{figure*}
  \includegraphics[width=1.7in,angle=-90]{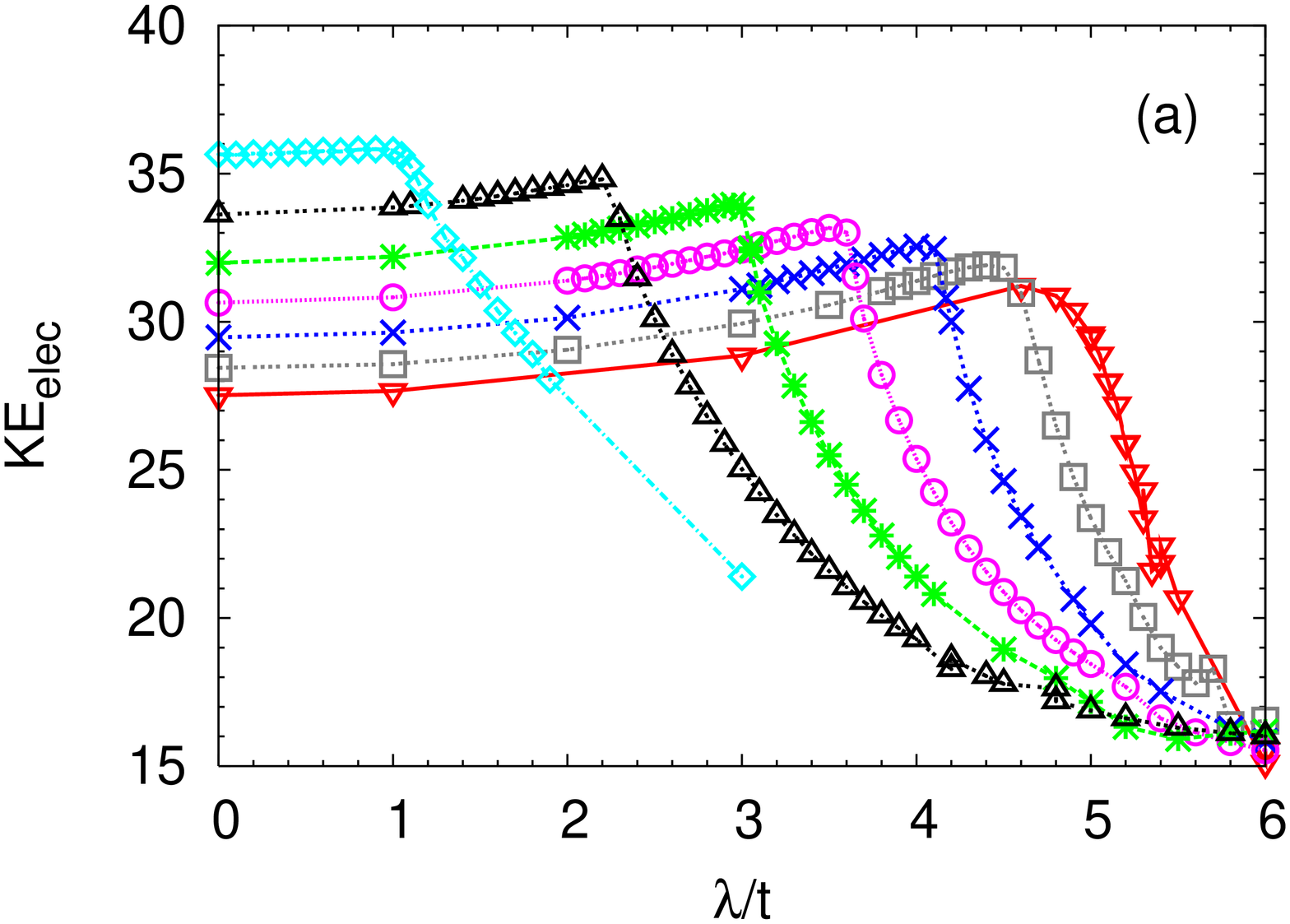}
  \includegraphics[width=1.7in,angle=-90]{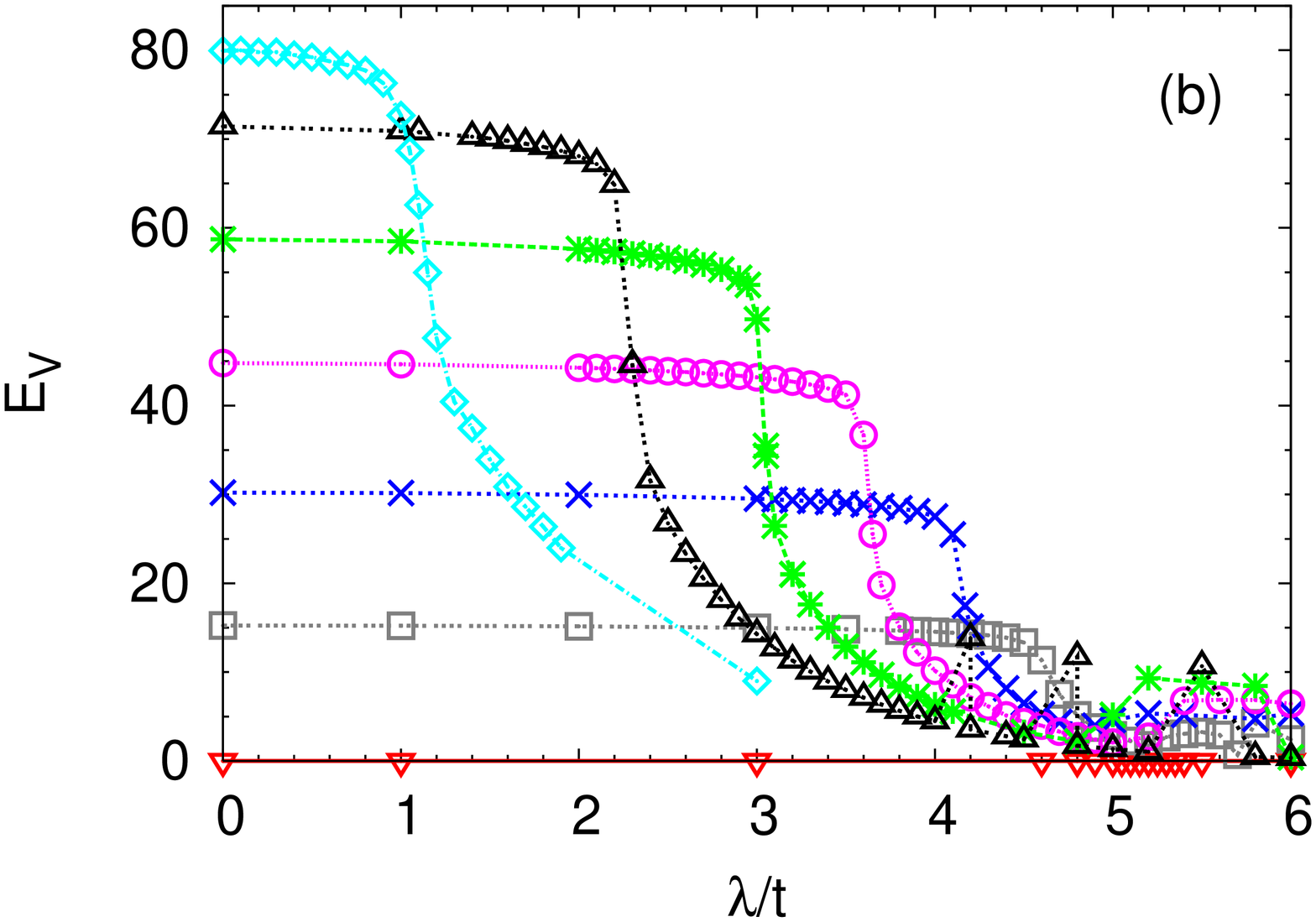}
  \includegraphics[width=1.7in,angle=-90]{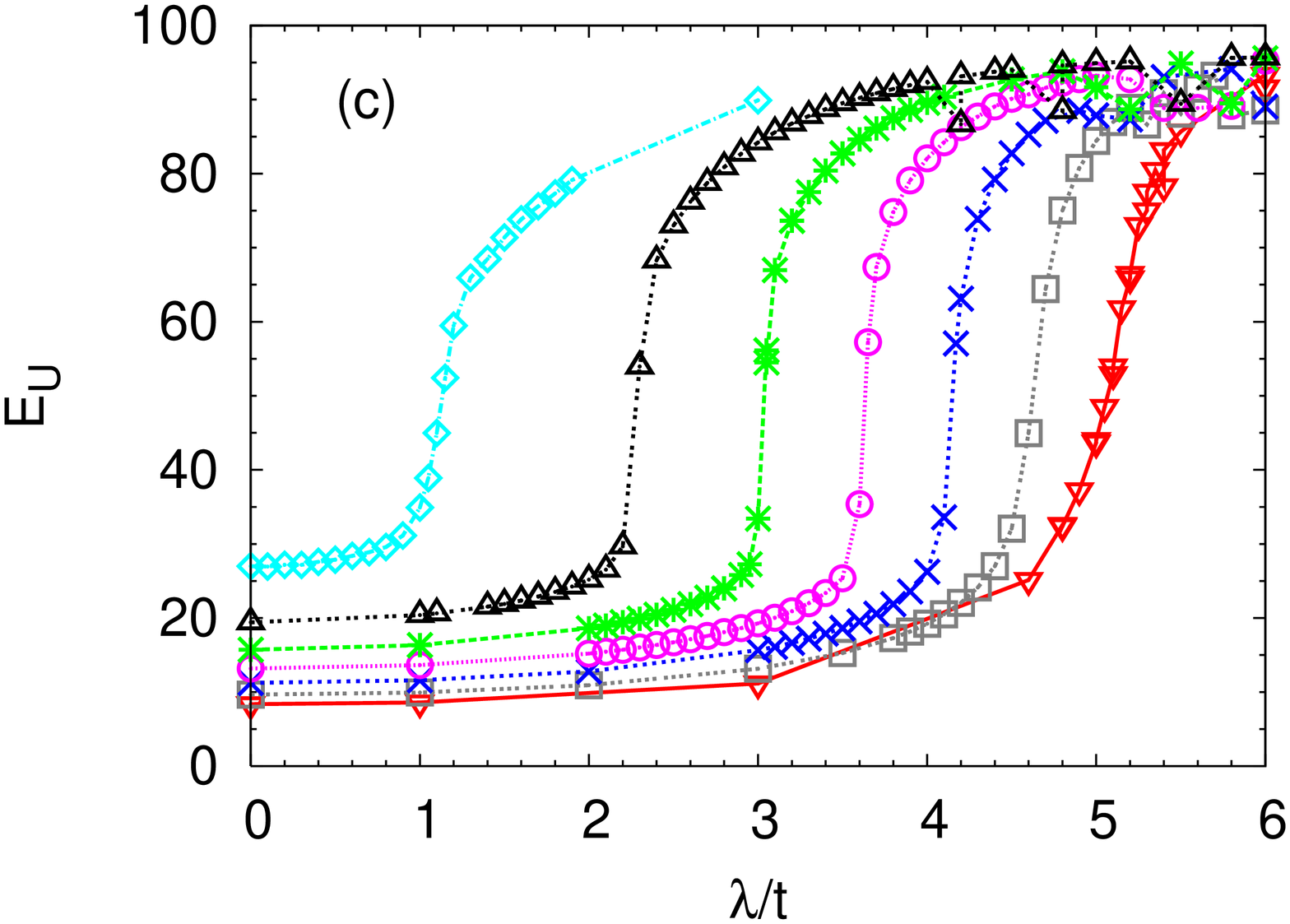}
  \includegraphics[width=1.7in,angle=-90]{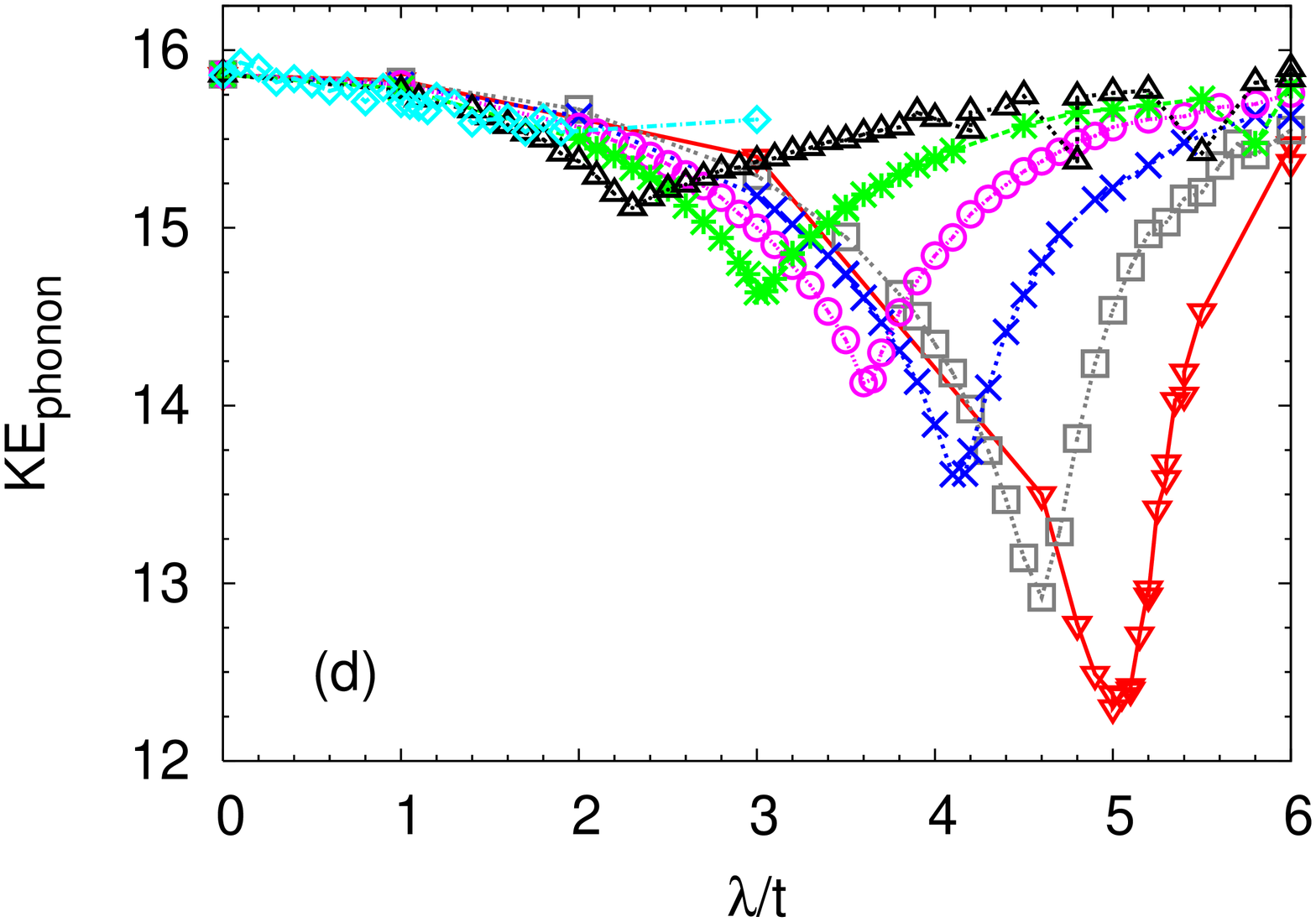}
  \includegraphics[width=1.7in,angle=-90]{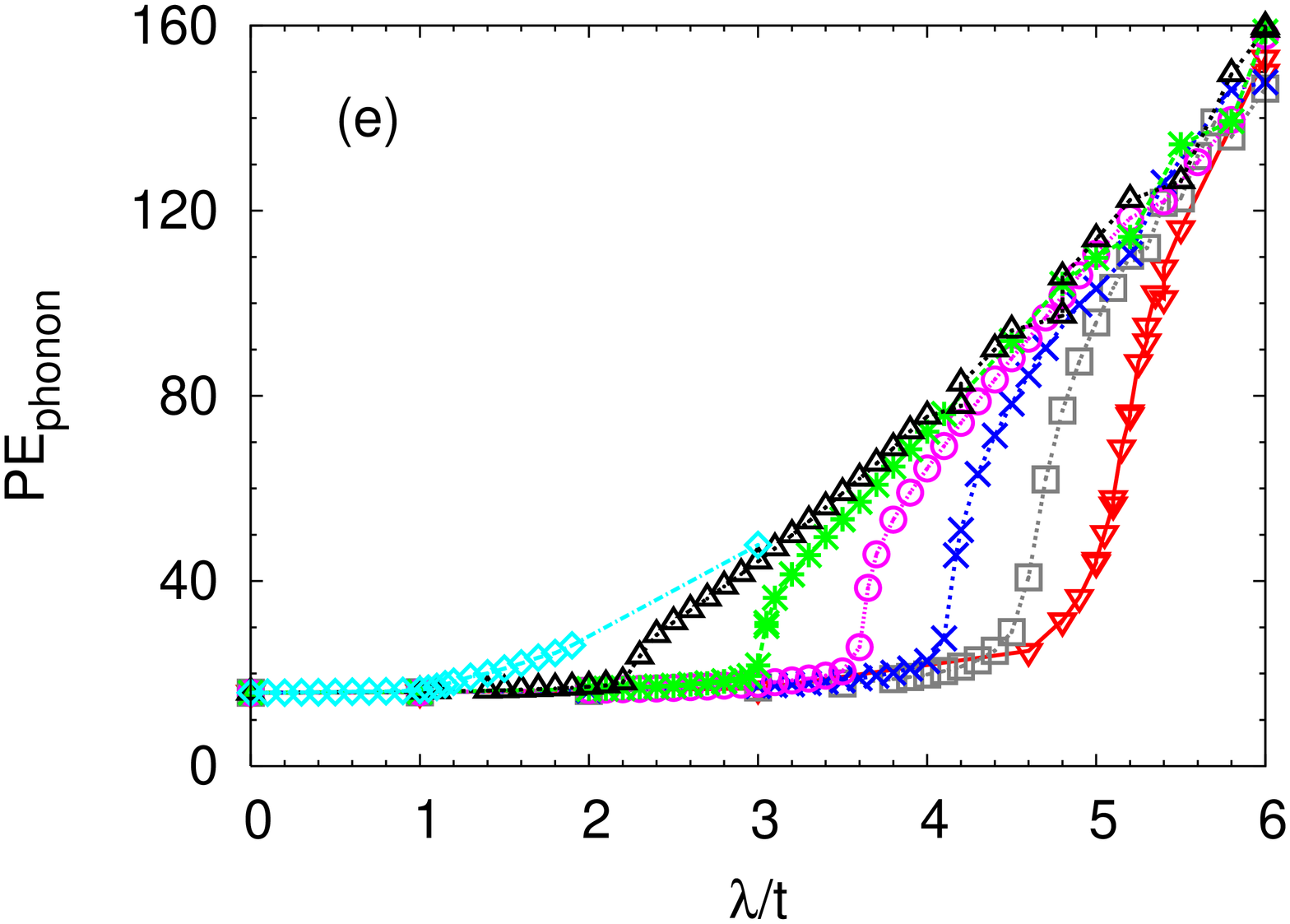}
  \includegraphics[width=1.7in,angle=-90]{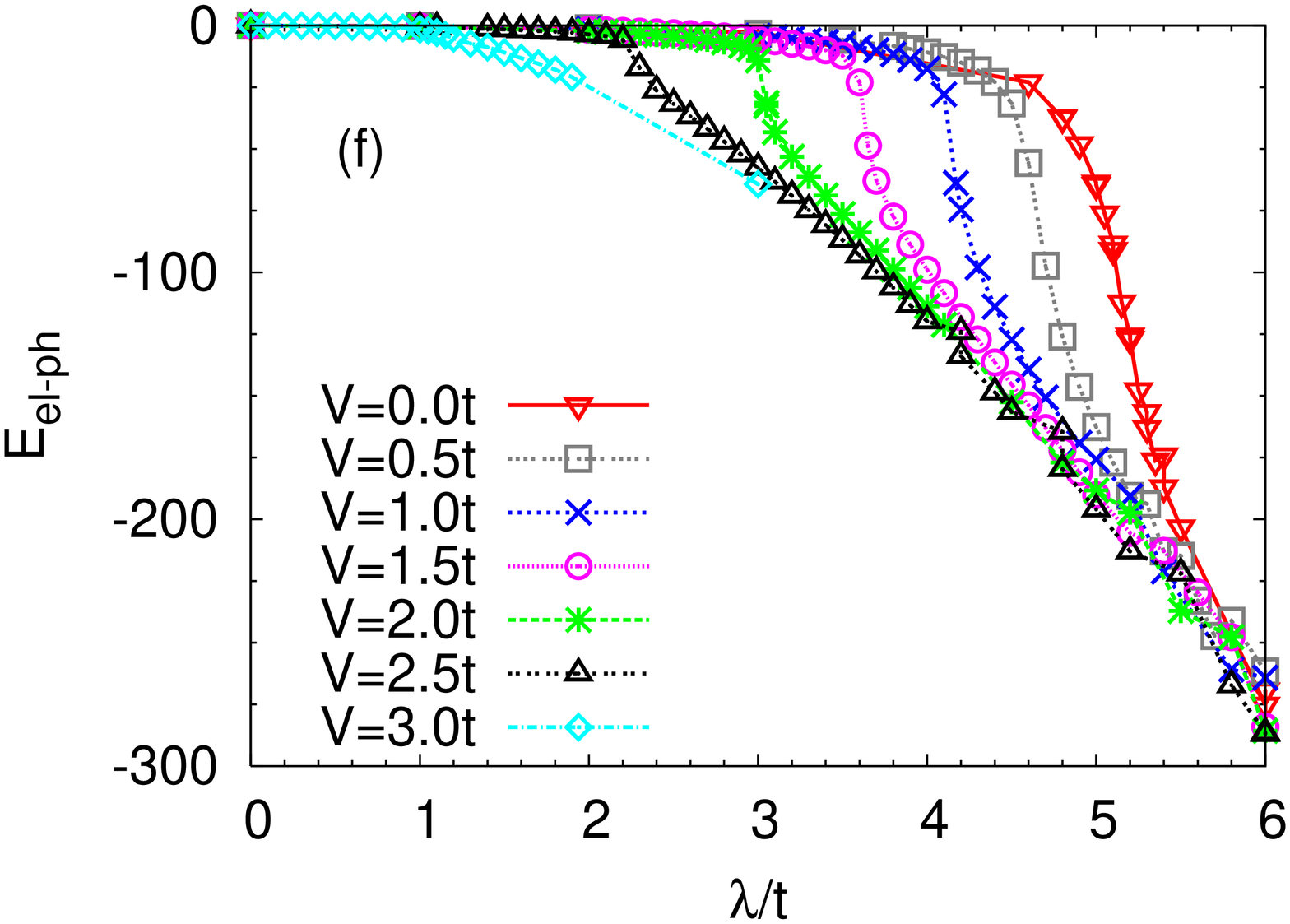}
  \includegraphics[width=1.7in,angle=-90]{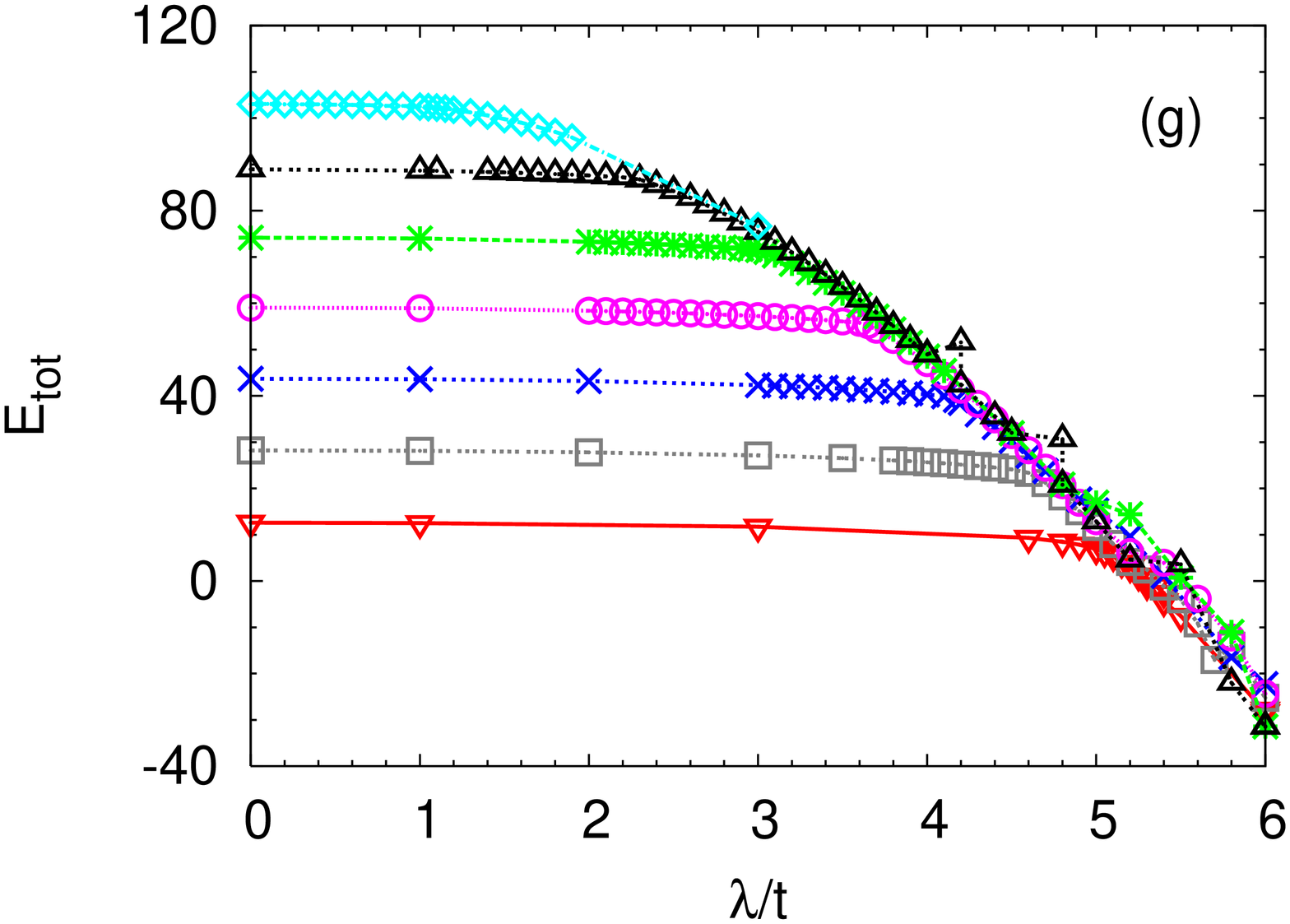}
  \caption{(a) Kinetic (hopping), (b) intersite Coulomb, (c) on-site
  Coulomb, (d) kinetic (phonon), (e) potential (phonon), (f)
  electron-phonon coupling, and (g) total energies versus coupling
  constant $\lambda$ for $U = 6t$, $V = 0.0t, (0.5t), 3.0t$, $\beta t
  = 8$, $\omega_0 = 2t$, and $N=32$.  The fluctuations in the phonon
  kinetic energy are significantly smaller than those for all other
  energies for these parameters.
 \label{fig:phn_en}}
\end{figure*}

In Fig.~\ref{fig:stg_susc} we fix $U=6t$ and the lattice size at
$N=32$, and sweep $\Delta$ for different choices of $V$.  As expected,
the size of $\Delta$ required to destroy the SDW phase decreases as
the intersite interaction $V$, which cooperates with $\Delta$, rises.
As with the data of Fig.~\ref{fig:stg_unscld_u6v1.5}, the fall of
$\chi_{\rm SDW}$ coincides closely with the rise of $\chi_{\rm CDW}$.
In each case the transition is marked also by a maximum in $\chi_{\rm
BOW}$.  The sharpness of the peak in $\chi_{\rm BOW}$ diminishes as
$V$ grows, which is consistent with the SSE\cite{sandvik03} and
density matrix (DM) RG\cite{jeckelmann02} calculations on the extended
Hubbard model which (although they disagree in certain respects) both
conclude that BO is not present at strong coupling.  We note that a
Mott-insulator--BO transition has also been suggested by Zhang {\it et
al.}  in the $V=0$ limit with $\Delta=2.0$ and $U_c = 5.95 \pm 0.01$.
\cite{zhang03}

The behavior of the total energy, Fig.~\ref{fig:stg_en}(a), is
featureless through the sweep upward in $\Delta$.  However, abrupt
evolution of the individual components of the energy,
Figs.~\ref{fig:stg_en}(b)-(d), accompanies the transitions in the
susceptibilities.  The energy associated with $V$ decreases sharply
upon exiting the SDW phase where adjacent sites are occupied, while
the energy associated with $U$ jumps upward with the development of
double occupancy.  The kinetic energy is relatively benign, but, like
$\chi_{\rm BOW}$, reaches maxima along the SDW-CDW transition line.
Evidently, the near balance between the insulating tendencies of $U$
and $V$ allows greater fluctuation in the electron positions.

The values of $\Delta$ at which the different susceptibilities change
abruptly, and at which features in the energy are also evident, enable
us to draw the phase diagram in the $V$-$\Delta$ plane for fixed $U=6t$
shown in Fig.~\ref{fig:stg_pd}.  At $\Delta=0$ our QMC results match
quite nicely the DMRG results of Jeckelmann,\cite{jeckelmann02} who
finds $V_c=3.155 \pm 0.005$ for $U=6t$.  This $\Delta=0$ transition
point is not too far shifted from the strong coupling value $V_c=U/2 =
3t$ when $U=6t$.

As the staggered potential $\Delta$ becomes greater, our QMC phase
boundary bends more away from the $t=0$ line $V_c = U/2 - \Delta$.
The SDW phase appears to terminate at $\Delta = 2.46 \pm 0.05$ in the
absence of intersite repulsion $V$.  While labeled as CDW, the large
$\Delta$ phase in the $V = 0$ limit is perhaps more properly termed a
band insulator, where the alternating charge density is a consequence
of the staggered one-body potential as opposed to many-body effects.

\begin{figure}[t]
  \includegraphics[width=2.0in,angle=-90]{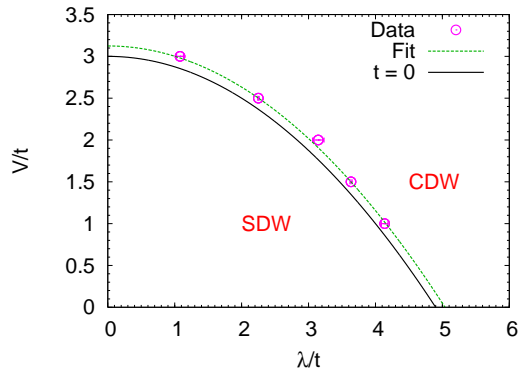}
  \caption{This figure shows the phase diagram for intersite $V$ and
  electron-phonon coupling $\lambda$ with $U = 6t$, $\omega_0 = 2t$,
  and $\beta t = 8$.  Line with symbols is the result of the WLQMC
  simulations in this paper.  The functional form of the fit is $V = a
  \lambda^2 + b$. We show the exact result (line without symbols) for
  the $t = 0$ phase.
    \label{fig:phn_pd}}
\end{figure}

\section{Results: Extended Hubbard Holstein Hamiltonian}
\label{sec:results_ehhh}

Having completed our discussion of the case of the interplay of a
static alternating external potential with the correlation terms $U,V$
in the extended Hubbard Hamiltonian, we now give analogous results for
the case when we couple to dynamical (Holstein) lattice distortions.
Figure \ref{fig:phn_unscld_u6v1.5} is a companion to
Fig.~\ref{fig:stg_unscld_u6v1.5}, showing the evolution of the
spin, charge, and bond susceptibilities with electron-phonon coupling
$\lambda$ (rather than staggered potential $\Delta$) for different
system sizes $N$.  As discussed earlier, $\lambda$ has a similar
qualitative effect to $\Delta$, since it weakens the on-site repulsion
$U$ and hence drives CDW formation.  There are significant
quantitative differences between the two situations.  The SDW-CDW
transition as a function of electron-phonon coupling $\lambda$ appears
to be much more abrupt.  Recall that $\Delta$ breaks the lattice
symmetry explicitly, selecting out a single preferred sublattice.  It
induces CDW order even within the SDW phase and as a consequence the
change through the transition is less dramatic.  The Holstein
interaction, in contrast, spontaneously breaks the translational
symmetry when it drives CDW order.  We note further that BOW order is
less sharply peaked at the SDW-CDW boundary.

Figure \ref{fig:phn_susc} is a companion to Fig.~\ref{fig:stg_susc},
similarly showing the susceptibilities as a function of
electron-phonon coupling constant $\lambda$ for a collection of values
of $V$ at a single lattice size $N=32$ and $\omega_0=2t$.  As $V$
increases, a smaller $\lambda$ is sufficient to drive CDW formation.
There appears to be some variation of the sharpness of the evolution
of the susceptibilities near $\lambda_c$ as $V$ is varied, with the
most abrupt behavior occurring for intermediate $V$.  In the extended
Hubbard model ($\lambda=0$), the transitions become monotonically more
steep with increasing $V$.  Indeed, as noted earlier, they change
from continuous to discontinuous beyond the tri- (multi)critical
point.  The fluctuations of $\chi_{\rm CDW}$ at large $\lambda$ in
Fig.~\ref{fig:phn_susc} (middle panel) often occur in QMC studies of
electron-phonon Hamiltonians and are associated with long
equilibration times which occur when the electrons and lattice degrees
of freedom are strongly coupled.

As with $\widehat H_{\rm IHM}$, the components of the energy
(Fig.~\ref{fig:phn_en}) lend important supporting evidence for the
locations of the transition points.  The behavior of $E_V$ and $E_U$
is the same as that observed previously in Fig.~\ref{fig:stg_en}, and
is more or less clear: in the SDW phase most sites are singly occupied
and there is a significant contribution to $E_V$, which then drops
abruptly in the CDW phase where doubly occupied and empty sites
alternate.  In contrast, $E_U$ is small in the SDW phase since sites
are singly occupied, but then increases sharply in the CDW phase.
What is perhaps less intuitive is the evolution of the phonon
contributions to the energy.  As $\lambda$ grows, the $t=0$ analysis
suggests a smooth quadratic increase, $E_{\rm phonon}^{\rm pot}=
\lambda^2 / 2\omega_0^2$.  Instead the phonon potential energy remains
relatively flat throughout the SDW region, and then jumps up as the
CDW is entered.  The phonon kinetic energy is especially interesting,
showing a well-defined {\it minimum} in the transition region.  The
origin of this effect is not clear.  $E_{\rm phonon}^{\rm kin}$ is
measured by the fluctuations of the phonon coordinates in imaginary
time.  Naively, one might expect kinetic lattice fluctuations to be
{\it largest} in the SDW-CDW transition region where the system is
undecided between which type of order to assume.  In the case of the
electron kinetic energy we see precisely this effect in
Fig.~\ref{fig:phn_en}(a).  The opposite appears to be the case for the
phonon kinetic energy.

Finally, Fig.~\ref{fig:phn_pd} shows the phase diagram in the
$V$-$\lambda$ plane at fixed $U=6t$.  It shares the same general
features as Fig.~\ref{fig:stg_pd} with a SDW phase near the origin
that is destroyed when either the intersite repulsion $V$ or the
electron phonon coupling $\lambda$ increases sufficiently.  Figure
\ref{fig:phn_pd} describes how large a value of electron-phonon
coupling $\lambda$ is required to convert the SDW phase, favored by
$U$, to the CDW phase, favored by $V$, and is representative of how
$\lambda$ affects the extended Hubbard model phase diagram at all
intermediate to large interaction strengths.  It is important to note
that, unlike Fig.~\ref{fig:stg_pd}, the QMC phase boundary does not
bend away from the $t=0$ line $V_c = U/2 - \lambda^2 / 4
\omega_0$. Instead, the boundary is uniformly shifted to increase the
critical intersite repulsion, favoring SDW order.  Again, the
$\lambda=0$ point on our phase boundary ($V_c = 3.124 \pm 0.011$)
agrees well with Jeckelmann's DMRG treatment.  (See above discussion
of Fig.~\ref{fig:stg_pd}.)

For finite $\lambda$ we can compare against the phase diagram of Sil
and Bhattacharyya who study the same extended Hubbard model coupled to
Holstein phonons. \cite{sil96} They draw the phase boundary in the
$U$-$V$ plane for different electron-phonon couplings.  Translating to
the units used in our paper, for $U=6t$ and $V=2t$, their data suggest
that the CDW phase is destroyed at $\lambda_c \approx 2.8$.  Our
Fig.~\ref{fig:phn_pd} gives $\lambda_c \approx 3.0$ at $V = 2t$ for
the same parameters.  Likewise, Sil and Bhattacharyya find that for
$\lambda=5.6$ there is no SDW phase at $U=3t$.  This is again nicely
consistent with our data, which suggest that when $\lambda = 5.04 \pm
0.06$ there is CDW order.

\section{Summary}
\label{sec:summary}

In this paper we have presented world-line quantum Monte Carlo
simulations of the extended one-dimensional Hubbard Hamiltonian to
which coupling to static staggered (ionic Hubbard) or dynamic
(Holstein) lattice degrees of freedom is added.  The evolution of the
susceptibilities to different types of order and the components of the
energy were examined.  For both static and dynamic couplings the
region of charge density wave order in the phase diagram is found to
be stabilized, and the phase boundaries are pinned down.  Bond order
is shown to be enhanced in the vicinity of the spin density to charge
density transition.

The results obtained in this work also show good agreement with
previous studies. The zero-coupling limit ($\Delta,\lambda=0$) results
conform well with Jeckelmann's DMRG results.  For dynamic couplings,
the results compare favorably with the results of Sil {\it et al}.

A comparison of the QMC phase boundaries with their counterpart in the
$t=0$ limit shows a markedly different behavior between the two types
of coupling. For static couplings, the CDW phase is enhanced in the
QMC calculation. Conversely, there is an enhancement of the SDW
correlations with Holstein phonons.

\vspace{20pt}
\section*{ACKNOWLEDGEMENTS}
\label{sec:acknowledgements}
\vspace{-12pt}
We acknowledge support from the DOE under Grant No. DE-FG01-06NA26204
and the NSF Grant No. REU PHY-0243904, and useful input from
G.K. Pips.

\bibliography{uvdelta9}

\begin{thebibliography}{56}
\bibitem[{not()}]{note1}
\bibinfo{note}{The model with random, rather than alternating, static
  distortions is often referred to as the binary alloy Hubbard Hamiltonian. The
  possibility of a Mott transition away from half filling has recently been
  discussed (see Ref. 15).}
\bibitem[{glo()}]{global}
\bibinfo{note}{In addition to moves that change a phonon coordinate at a single
  space-time point, we also include moves that modify the phonon coordinates at
  all imaginary times for a given spatial site.}
\end{thebibliography}

\end{document}